\documentclass[]{aastex631}

\usepackage{mathptmx}
\usepackage{amsmath}
\usepackage{threeparttable}
\usepackage{xcolor}
\usepackage{booktabs}
\usepackage{graphicx}
\usepackage{rotating}
\begin{document}

\title{A Catalog of 486 Barium Stars Identified in GALAH DR4}

\author{Guochao Yang}
\altaffiliation{Contact e-mail: gcyang@nao.cas.cn}
\affiliation{School of Physics and Astronomy, China West Normal University, 637009 Nanchong, People's Republic of China}
\affiliation{National Astronomical Observatories, Chinese Academy of Sciences, 100012 Beijing, People's Republic of China}
\author{Jingkun Zhao}
\altaffiliation{Contact e-mail: zjk@nao.cas.cn}
\affiliation{National Astronomical Observatories, Chinese Academy of Sciences, 100012 Beijing, People's Republic of China}
\author{Nian Liu}
\altaffiliation{Contact e-mail: nian@cwnu.edu.cn}
\affiliation{School of Physics and Astronomy, China West Normal University, 637009 Nanchong, People's Republic of China}
\author{Xiaokun Hou}
\affiliation{National Astronomical Observatories, Chinese Academy of Sciences, 100012 Beijing, People's Republic of China}
\author{Zhenxin Lei}
\affiliation{Physics Department, Xiangtan University, 411105 Xiangtan, People's Republic of China}
\author{Xianhao Ye}
\affiliation{National Astronomical Observatories, Chinese Academy of Sciences, 100012 Beijing, People's Republic of China}
\affiliation{Instituto de Astrof\'{i}sica de Canarias, V\'{i}a L\'{a}ctea, 38205 La Laguna, Tenerife, Spain}
\affiliation{Universidad de La Laguna, Departamento de Astrof\'{i}sica, 38206 La Laguna, Tenerife, Spain}
\author{Gang Zhao}
\affiliation{National Astronomical Observatories, Chinese Academy of Sciences, 100012 Beijing, People's Republic of China}
\affiliation{Institute of Space Sciences, School of Space Science and Technology, Shandong University, 264209 Weihai, People's Republic of China}
\begin{abstract}
We present a catalog of 486 Ba stars identified in the GALAH~DR4 survey using high-resolution
spectra and precise abundance measurements. The sample was selected based on $s$-process enrichment
criteria involving the abundances of Ba and La relative to Eu, and further refined using the
signed-distance method, resulting in the largest sample of Ba stars to date, including five newly
identified Ba dwarfs. Using astrometric and kinematic parameters from Gaia~DR3 and StarHorse,
we derived the Galactic velocity components for 367 Ba stars and found that most belong to the
thin or thick disk, while 18 exhibit halo-like kinematics. Chemical abundance analysis suggests
that most Ba stars are of in situ origin, whereas two stars (4077588766331013248 and
6692980582560946304) display signatures of accreted populations. The $E$–$L_{\rm z}$ diagram
further shows that star 4077588766331013248 lies within the region of substructure ED-8.
The observed decline of [Ba/Fe] and [La/Fe] with increasing metallicity implies that $s$-process
elements originate from nucleosynthetic sites distinct from those of iron-peak elements, while
the decreasing [hs/ls] ratio toward higher metallicity indicates higher neutron-capture efficiency
at lower metallicity. Both giant and dwarf Ba stars—--except for 6053735173729807872—--are
confirmed binaries hosting white dwarf companions. Moreover, we estimated the masses of the
former asymptotic giant branch (AGB) companions to these Ba stars based on the good agreement
between the FRUITY AGB model predictions and their observed abundance patterns.
\end{abstract}


\section{Introduction}

Barium (Ba) stars have long been a subject of considerable interest in stellar
astrophysics, as they constitute a class of chemically peculiar giants and dwarfs
characterized by overabundances of slow neutron-capture ($s$-process) elements such
as Ba, La, Ce, and Nd \citep{Bidelman1951,Burbidge1957}. They typically
span spectral types from G to K, with effective temperatures of $~$4000--6000~K.
The peculiar chemical patterns of these stars have long attracted attention, as
they cannot be explained by standard stellar evolution. It is widely accepted
that $s$-process elements are synthesized in low- and intermediate-mass
stars with $M \sim 1.3$--$8~M_{\odot}$ during the asymptotic giant branch
(AGB) phase, following the red giant stage \citep{Gallino1998,Busso1999,kappeler2011}.
Early studies of radial-velocity variations revealed that many Ba stars are
members of binary systems, with unseen white dwarf (WD) companions that were
formerly AGB stars \citep{McClure1980,McClure1990,Han1995,Jorissen1998,Jorissen2019}.
Mass transfer from these companions is believed to be the primary mechanism
for the observed $s$-process enrichment, highlighting the close connection between
chemical peculiarities, AGB stars, and binary evolution \citep{Cseh2018,Roriz2021a,
Roriz2021b,Stancliffe2021,Goswami2023,Roriz2024}.

Historically, the study of Ba stars has progressed from initial spectroscopic
identifications \citep[e.g.,][]{Warner1965,Lu1983,Lu1991} to detailed abundance
analyses \citep[e.g.,][]{Zacs1994,Liang2000,Liang2003,Liu2009}. Systematic
catalogs and high-resolution spectroscopic observations have allowed researchers to
investigate correlations between stellar atmospheric parameters, elemental
abundances, and companion masses \citep{Allen2006a,Allen2006b,Pereira2011,
deCastro2016,Yang2016,Karinkuzhi2018,Yang2024}. These studies showed that the
degree of $s$-process enhancement is consistent with nucleosynthesis in AGB stars
and depends on the properties of the binary system, such as orbital separation
and companion mass. In addition, variations in the ratios of light (e.g., Sr, Y,
and Zr) to heavy (e.g., Ba, La, Ce, and Nd) $s$-process elements provide insights
into the efficiency of neutron-capture processes and material accretion, offering
a valuable probe of AGB nucleosynthesis and binary interaction mechanisms
\citep{deCastro2016,Cseh2018,Roriz2021a,Roriz2021b}.

Compared to the relatively large samples of Ba giants analyzed through
high-resolution spectroscopy—--for instance, 169 and 180 stars reported by
\citet{deCastro2016} and \citet{Roriz2021a}, respectively—--the number of Ba
dwarfs confirmed from detailed chemical analyses remains small, with only
about 71 identified to date \citep[e.g.,][]{Kong2018,Purandardas2019,
Shejeelammal2020,Liu2021,Karinkuzhi2021b,Roriz2024}. Despite decades of
observational and theoretical efforts, the total number of Ba stars in the
Galaxy remains uncertain. \citet{Han1995} estimated that up to ~22,000 Ba stars
with $V \leq 10.0$ may exist in the Milky Way. Thanks to the extensive datasets
delivered by the large-scale spectroscopic surveys such as the Large Sky
Area Multi-Object Fiber Spectroscopic Telescope \citep[LAMOST;][]{Cui2012,Zhao2012}
and Galactic Archeology with HERMES \citep[GALAH;][]{DeSilva2015,
Sheinis2015,Bunder2024}, it has become feasible to conduct systematic and
large-scale searches for Ba stars \citep{Norfolk2019,Rekhi2024,Chen2025}.
Moreover, detailed chemical abundance patterns, particularly for
neutron-capture elements, show substantial diversity. Since the nucleosynthetic
processes operating in AGB stars are not yet fully understood, large
homogeneous spectroscopic surveys now provide an essential means to investigate
Ba stars systematically, increasing both sample sizes and the precision of
abundance measurements.

In this study, we present a comprehensive analysis of Ba stars based on the
fourth data release (DR4) of the GALAH survey \citep{Bunder2024}, providing a
catalog of 486 stars with well-determined stellar parameters. We examined
their elemental abundance patterns, focusing on $s$-process enrichment,
and investigate their kinematic properties to explore connections between
chemical peculiarities and Galactic dynamics. Our analysis aims to provide
new insights into the formation history, binary evolution, and chemical
characteristics of Ba stars. The structure of this paper is organized as
follows. Section 2 outlines the data selection and methodology. Section 3
presents the analysis of the kinematic properties, progenitor systems,
chemical abundances, and binarity of the Ba stars. Finally, Section 4
summarizes the main conclusions.

\section{Data and methods}

In this study, we utilized data from the GALAH~DR4 survey \citep{Bunder2024},
which provides high-resolution (R $\sim$ 28,000) optical spectra for 917,588
stars observed with the HERMES spectrograph on the 3.9~m Anglo-Australian
Telescope. The catalog delivers homogeneous estimates of stellar atmospheric
parameters ($T_{\rm eff}$, $\log g$, [Fe/H], and $\xi_{\rm t}$)
along with abundances for up to 32 chemical species. Such a large
and uniform data set offers unprecedented opportunities to probe the chemical
enrichment and dynamical history of the Milky Way, while also informing broader
investigations into the origin of elements and the stellar evolution. To
enhance our analysis, we complemented the GALAH spectroscopy with high-precision
astrometry from Gaia DR3 \citep{Gaia2023} and distance constraints from the
StarHorse catalog \citep{Queiroz2023}.

We adopted the standard convention for elemental abundances. For an element $i$,
its absolute abundance is defined as $A(\rm i) = \log_{10}(N_{i}/N_{\rm H}) + 12$,
where $N_{\rm i}$ and $N_{\rm H}$ are the number densities of element $i$ and hydrogen,
respectively. Based on the measured abundance ratios $[{\rm i}/{\rm Fe}]$
and $[{\rm Fe}/{\rm H}]$, we derived the absolute abundance as
$A_{\rm i,*} = [{\rm i}/{\rm Fe}] + [{\rm Fe}/{\rm H}] + A_{\rm i,\odot}$,
where $A_{\rm i,\odot}$ is the adopted solar photospheric abundance from \citet{Asplund2009}.

To ensure the reliability of our sample, we first applied quality flags in GALAH DR4 to
exclude stars with unreliable stellar parameters or elemental abundances. Ba and La
are taken as representative $s$-process elements, whereas Eu is considered a typical
$r$-process element, primarily produced in core-collapse supernovae or neutron star
mergers \citep{Beers2005,Qian2007,Abbott2017,Stancliffe2021}. Following the
classification scheme of \citet{Beers2005}, we retained stars with [Ba/Eu] $>$ 0.5
and [La/Eu] $>$ 0.5, indicative of $s$-process enrichment. Subsequently, candidate
Ba stars were identified using the [s/Fe] ratio, defined following the criteria of
\citet{deCastro2016}. The [s/Fe] ratio, defined as the mean abundance of several
main $s$-process elements---including not only Ba and La but also Sr, Y, Zr, Ce, and
Nd---relative to Fe \citep{deCastro2016}, serves as a primary diagnostic for $s$-process
enrichment. Stars with [s/Fe] values significantly higher than typical field stars
are classified as $s$-rich. Various thresholds have been proposed, ranging from
[s/Fe] $= 0.34$ \citep{Rojas2013} down to 0.25 \citep{deCastro2016}. In this
work, we adopted the latter threshold, consistent with \citet{Yang2024}, as the
first-step selection criterion for Ba star candidates.

\begin{figure}
\centering
\includegraphics[width=16cm]{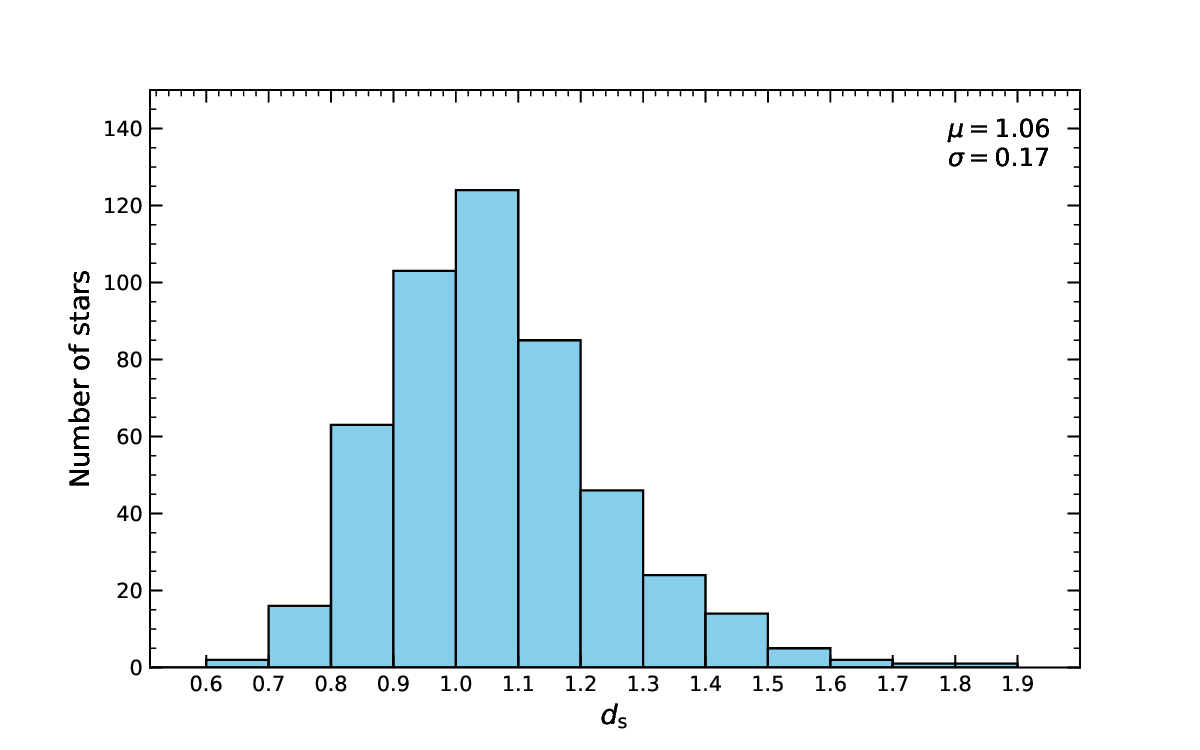}
\caption{Distribution of the $d_{\rm s}$ distances for the Ba stars in our sample.
The corresponding mean ($\mu$) and standard deviation ($\sigma$) are indicated in
the figure.}
\label{fig:ds-n}
\end{figure}

As a second-step verification, we employed the methodology of \citet{Karinkuzhi2021a},
who suggested a signed distance, $d_{\rm s}$, to quantify the deviation of a star's
abundance pattern from the solar $r$-process composition:
\begin{equation}\label{dsequa}
d_{\rm s} = \frac{1}{N}\sum_{i=1}^{N} \left(A_{\rm i,\star} - A_{\rm i,r,\odot}\right),
\end{equation}
where $A_{\rm i,r,\odot}$ is the solar $r$-process abundance from \citet{Goriely1999},
scaled to match the Eu abundance of the target star. The element set considered
is $i = {\rm Sr, Y, Zr, Ba, La, Ce, Nd}$. In this scheme, $d_{\rm s}$ increases with
higher $s$-process enrichment, reflecting a stronger deviation from the solar $r$-process
pattern. Following \citet{Karinkuzhi2021a}, we adopted $d_{\rm s} = 0.6$ as the threshold
for Ba stars and used this criterion to perform a second-step verification of the
selected candidates. As a result, the number of Ba star candidates decreased from
488 to 486, with two stars (3419017310212847488 and 3308913761994892672) showing
[s/Fe] $> 0.25$ but lacking reliable $d_{\rm s}$ values due to unreliable [Fe/H]
determinations. The distribution of $d_{\rm s}$ for the Ba stars is shown in
Figure~\ref{fig:ds-n}. As illustrated in the figure, the $d_{\rm s}$ values of the
sample stars range from 0.68 to 1.83, with the distribution peaking at a
mean value of $\mu$ = 1.06 and a standard deviation of $\sigma$ = 0.17.

Based on the computation and selection criteria described above, we constructed
a catalog of Ba stars from GALAH DR4. The final catalog contains 486 Ba stars,
as listed in Table~\ref{tab:catlog}. In this table, stellar identifiers
from GALAH~DR4, Two Micron All Sky Survey (2MASS), and Gaia~DR3 are all provided.
Throughout the paper, we adopt the Gaia~DR3 source IDs when referring to specific
objects. The R.A., decl., stellar masses, atmospheric parameters,
and heavy-element abundances are taken from GALAH~DR4. The $V_{JKs}$ magnitudes
of the sample stars are computed using Equation (1) of \citet{Bunder2024},
based on the 2MASS $J$ and $K_{\rm s}$ magnitudes. The diagnostic indicators of
$s$-process enrichment, [s/Fe] and $d_{\rm s}$, as well as the inferred AGB companion
masses of Ba stars, $M_{\rm AGB}$, are derived in this work and discussed in the
following sections.

\begin{figure}
\centering
\includegraphics[width=16cm]{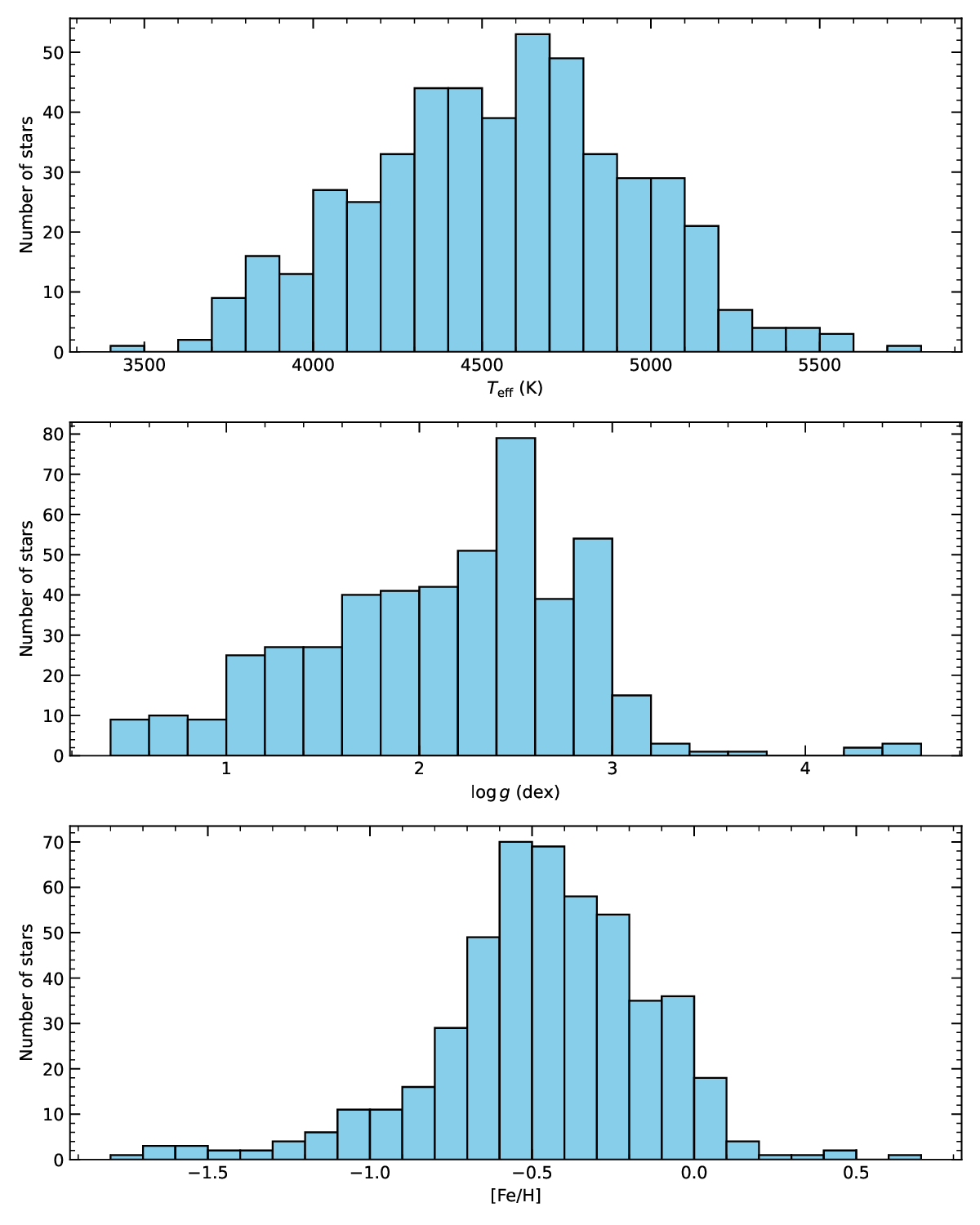}
\caption{Histograms of $T_{\rm eff}$, $\log g$, and [Fe/H] for the Ba star sample.}
\label{fig:parameter}
\end{figure}

For a clearer view of the distributions of the stellar atmospheric parameters,
$T_{\rm eff}$, $\log g$, and [Fe/H], we present in Figure~\ref{fig:parameter} the
corresponding histograms for the final Ba star sample. As shown in the figure,
the sample mainly spans effective temperatures of $\sim$3500--5500~K and surface gravities
from $\log g \approx 1.0$ to 3.0~dex, indicating that the majority of the stars are giants
and subgiants. The metallicity distribution peaks around [Fe/H] $\approx -0.5$, with
a tail extending toward lower metallicities, reflecting the metal-poor nature of a
significant fraction of the Ba stars. These distributions provide a global overview
of the parameter space covered by our sample.

\begin{figure}
\centering
\includegraphics[width=16cm]{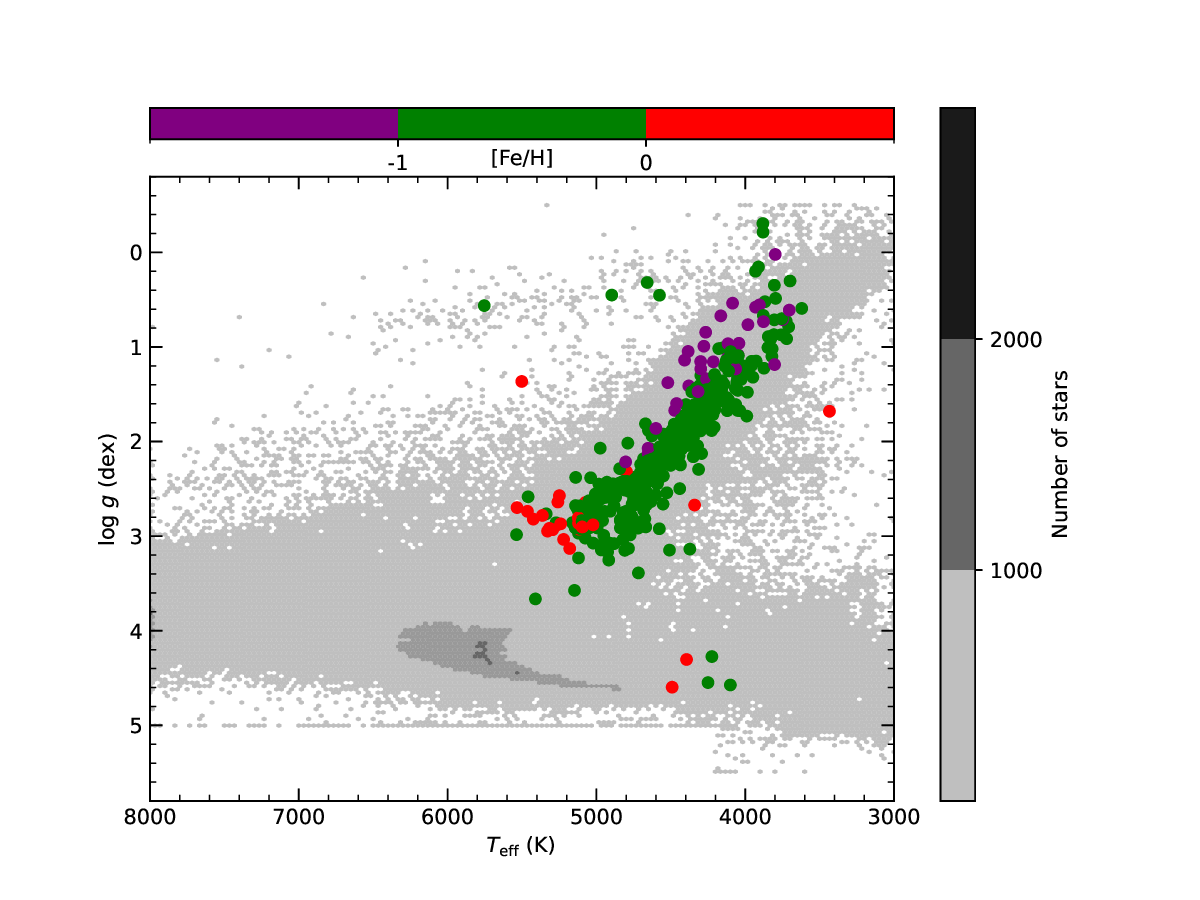}
\caption{Kiel diagram of Ba stars in GALAH, with points colored according to
their metallicities. For reference, all GALAH DR4 stars are plotted in the
background as filled gray circles.}
\label{fig:hr}
\end{figure}

The distribution of the sample stars in the effective temperature
versus surface gravity diagram is shown in Figure~\ref{fig:hr}. The majority of the Ba
stars are concentrated along the giant branch with $\log g < 4.0$~dex. We identify 481 Ba
giants out of a total of 475,914 giants in GALAH~DR4, corresponding to a fraction of
$\sim 0.1\%$. This value is lower than the commonly quoted incidence of $\sim 1\%$ for
Ba stars among giant populations reported in previous studies
\citep{Williams1975,Lu1991,Lebzelter2013}. The discrepancy is
primarily due to our conservative selection strategy. To ensure the reliability of the
final sample, we applied strict quality flags in GALAH~DR4 to exclude stars with unreliable
stellar parameters or elemental abundances, which significantly reduces the effective
parent sample. In addition, the combined requirements on [s/Fe] and the signed distance
$d_{\rm s}$ further restrict the sample to objects with robust $s$-process enrichment signatures.
As a result, our catalog represents a highly reliable, though incomplete, sample of Ba giants.

Five objects lie on the main sequence and are classified as Ba dwarfs with
$\log g > 4.0$~dex. These newly identified dwarfs---4904639351072483840,
4644665049364988928, 3220808727029495680, 5480510146667329152, and 6810035273351899648
---increase the known Ba dwarf population from 71 to 76 \citep{Roriz2024}. Despite the large
overall Ba star sample identified in GALAH~DR4, the number of Ba dwarfs remains strikingly
small compared to that of Ba giants. This imbalance is likely driven by strong observational
and evolutionary biases. As discussed by \citet{Kong2018} and \citet{Escorza2017,Escorza2019},
Ba dwarfs are intrinsically more difficult to identify because their higher effective
temperatures lead to weaker $s$-process absorption features, making chemical peculiarities
less conspicuous than in cooler giants. In addition, the relatively shallow convective
envelopes of main-sequence stars can dilute the accreted $s$-process material, further
reducing observable abundance enhancements. Classical Ba star surveys have also
historically focused on luminous G--K giants, introducing a selection bias against dwarfs.
Moreover, mass transfer via wind accretion may not yet have occurred or may not always
produce immediately detectable chemical signatures during the main-sequence phase,
particularly in long-period systems \citep{Escorza2019}.

\begin{table}[p]
\centering
\rotatebox{90}{
\begin{minipage}[c]{1.30\textheight}
\scriptsize
\setlength\tabcolsep{0.8pt}
\begin{threeparttable}
\caption{Properties of the 486 Ba stars.}
\label{tab:catlog}
\begin{tabular}{lcccccccccccccccccccccccc}
\toprule
GALAH ID & 2MASS ID & Gaia DR3 ID & R.A. & Decl & $V_{\rm JKs}$ & $M_{\rm Ba}$ & $T_{\rm eff}$  & $\log g$ & [Fe/H] &
$\xi_{\rm t}$ & [C/Fe] & [Sr/Fe] & [Y/Fe] & [Zr/Fe] & [Ba/Fe] & [La/Fe] & [Ce/Fe] & [Nd/Fe] & [Sm/Fe] & [Eu/Fe] &
[s/Fe] & $d_{\rm s}$ & $M_{\rm AGB}$ \\
  &  &  & (J2016.0) & (J2016.0) & (mag) & ($M_{\odot}$) & (K) & (dex) &  & (km\,s$^{-1}$) & & & & & & & & & & & & & ($M_{\odot}$) \\
\midrule
210719001601247 & 00112191-6201326 & 4904639351072483840 & 00:11:21.97 & -62:01:32.50 & 13.96 & 0.72 & 4490.90 & 4.60 & 0.04 & 0.66 &
-- & 0.58 & 0.33 & 0.08 & 0.67 & 0.92 & 0.78 & 0.57 & 0.33 & -0.02 & 0.56 & 1.12 & 1.5 \\
200905003101166 & 00150916-6657084 & 4707537593846950528 & 00:15:09.16 & -66:57:08.61 & 12.32 & 0.95 & 4380.84 & 2.04 & -0.53 & 1.52 &
-0.03 & 0.05 & 0.38 & 0.48 & 1.02 & 0.93 & 0.41 & 0.67 & 0.76 & 0.19 & 0.56 & 0.97 & 3.0  \\
201005003601167 & 00303605-6922184 & 4702688545706197760 & 00:30:36.10 & -69:22:18.41 & 12.85 & 0.99 & 4527.02 & 2.04 & -0.69 & 1.39 &
0.2 & --  & 0.58 & 0.91 & 1.35 & 1.37 & 1.02 & 1.09 & 0.68 & 0.52 & 1.05 & 1.02 & 1.5 \\
161116002201312 & 00365219-7311276 & 4688891117917292544 & 00:36:52.19 & -73:11:27.69 & 13.94 & 3.47 & 4298.13 & 1.30 & -0.57 & 2.65 &
0.15 & 0.16 & 0.13 & 0.17 & 0.58 & 0.89 & -0.36 & 0.29 & 0.13 & 0.00 & 0.27 & 0.81 & 1.3 \\
140814006001283 & 00464733-7431422 & 4685489538152953856 & 00:46:47.34 & -74:31:42.09 & 13.48 & 4.01 & 3867.45 & 0.52 & -0.35 & 2.69 &
-- & -0.09 & -0.24 & 0.04 & 0.94 & 0.48 & 0.34 & 0.54 & 0.29 & -0.03 & 0.29 & 0.88 & 3.0 \\
140807005601363 & 01112540-7317042 & 4687174196135131008 & 01:11:25.40 & -73:17:04.31 & 13.79 & 1.95 & 4177.95 & 1.02 & -0.79 & 2.73 &
0.21 & --  & -0.14 & -0.13 & 0.92 & 0.84 & 0.60 & 0.62 & 0.47 & 0.28 & 0.45 & 0.71 & 3.0 \\
151109002101111 & 01300971+0528538 & 2564906606055659520 & 01:30:09.71 & +05:28:53.80 & 12.30 & 1.09 & 4895.70 & 2.45 & -0.59 & 1.69 &
0.21 & -- &  0.67 & 0.30 & 1.25 & 1.23 & 1.00 & 0.94 & 0.56 & 0.48 & 0.90 & 0.91 & 1.5 \\
201001003201189 & 02074101-6458483 & 4699880461727404544 & 02:07:41.01 & -64:58:48.40 & 12.62 & 1.06 & 4838.88 & 2.91 & -0.29 & 1.23 &
0.05 & 0.18 & 0.55 & 0.32 & 0.83 & 0.66 & 0.62 & 0.41 & 0.51 & -0.02 & 0.51 & 1.10 & 1.5 \\
191106003301258 & 02124334-7323054 & 4644364539093274496 & 02:12:43.38 & -73:23:05.40 & 13.69 & 0.97 & 4683.97 & 2.52 & -0.48 & 1.31 &
0.15 & --  & 0.68 & 0.33 & 1.31 & 1.23 & 0.89 & 0.93 & 1.00 & 0.48 & 0.90 & 0.96 & 3.0 \\
140811005601156 & 02134217-7226579 & 4644665049364988928 & 02:13:42.21 & -72:26:57.90 & 13.94 & 0.65 & 4250.07 & 4.55 & -0.35 & 0.78 &
-- & --  & 0.17 &  -- & 0.27 & 1.07 & -- & -- & 0.13 & -0.25 & 0.50 & 1.17 & 1.5 \\
... & ... & ... & ... & ... & ... & ... & ... & ... & ... & ... & ... & ... & ... & ... & ... & ... & ... & ... & ... & ... & ... & ...  & ... \\
... & ... & ... & ... & ... & ... & ... & ... & ... & ... & ... & ... & ... & ... & ... & ... & ... & ... & ... & ... & ... & ... & ...  & ... \\
\bottomrule
\end{tabular}
\begin{tablenotes}
\scriptsize
\centering
\item Note. The star identifiers and the data in columns 4, 5, and 7–21 are taken from the GALAH survey.
\item (The complete table is available in machine-readable format.)
\end{tablenotes}
\end{threeparttable}
\end{minipage}}
\end{table}

\section{Analysis}
\subsection{Kinematic properties}

The kinematic analysis of stars in the solar neighborhood is commonly
performed in terms of the three Galactic velocity components $U$, $V$,
and $W$, which are defined as pointing toward the Galactic center,
along the direction of Galactic rotation, and toward the north Galactic
pole, respectively. Using precise parallaxes, proper motions, and
radial velocities from Gaia~DR3 \citep{Gaia2023}, we computed
the full three-dimensional space motions of our sample stars.

The heliocentric space velocities were computed following the
prescription of \citet{Johnson1987} and subsequently transformed to
the local standard of rest (LSR) using the solar motion vector
($U_\odot$, $V_\odot$, $W_\odot$) = (11.1, 12.24, 7.25) km\,s$^{-1}$
\citep{Schonrich2010}. To ensure the reliability of our sample,
we applied the following quality cuts prior to the velocity
calculations: (i) stars with a relative parallax error larger
than 20\% (parallax error/parallax $>$ 0.2) are discarded \citep{Carrillo2024};
(ii) distances were computed as the inverse of the Gaia
parallaxes whenever available and reliable. For stars with distances
exceeding 4~kpc, we adopted StarHorse parallaxes instead of Gaia
data; stars lacking both reliable Gaia and StarHorse parallaxes
were excluded to maintain the integrity of the results. After these
selections, space velocities were derived for a final sample of
367 Ba stars.

\begin{figure}
\centering
\includegraphics[width=14cm]{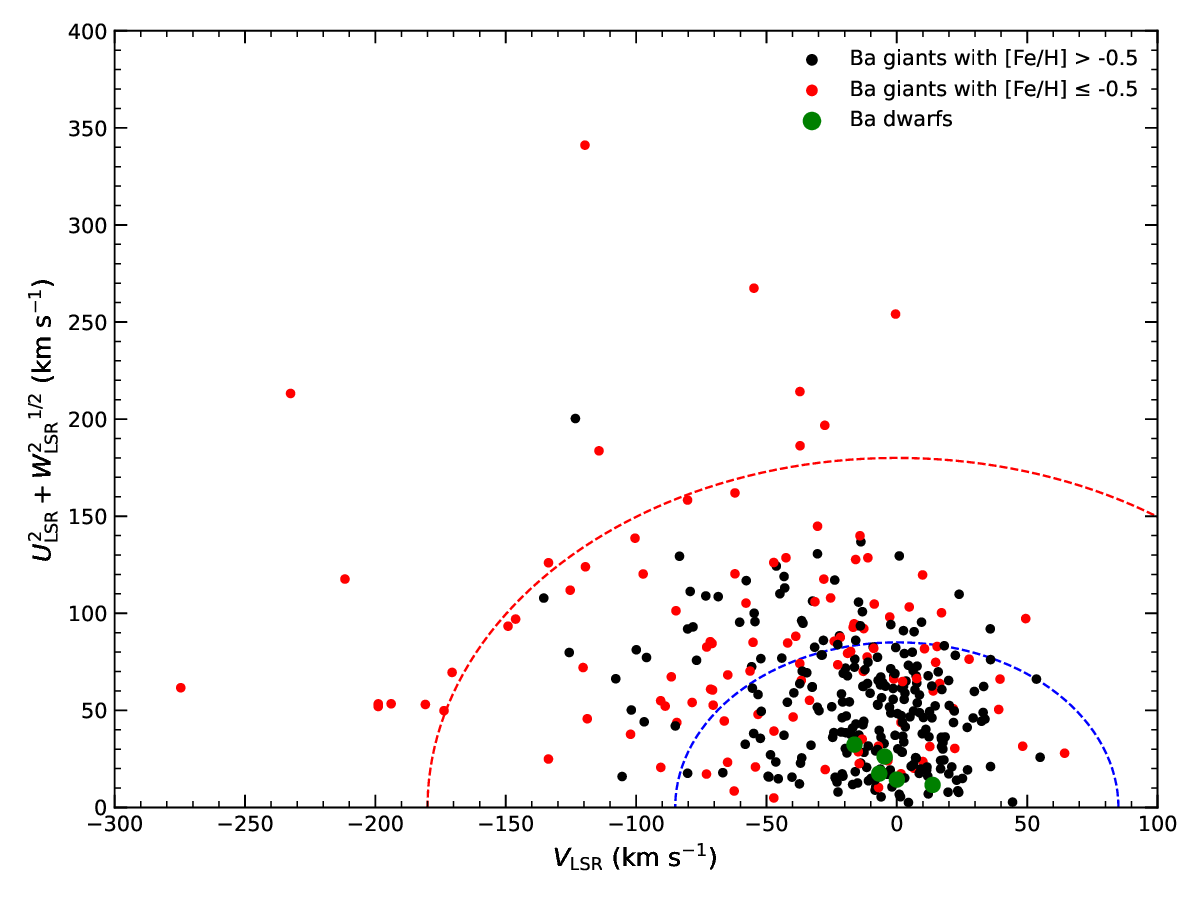}
\caption{The Toomre diagram of $(U_{\rm LSR}^{2}+W_{\rm LSR}^{2})^{1/2}$ vs.
$V_{\rm LSR}$ for our sample stars. Black- and red-filled circles represent
Ba giants with near-solar metallicities ($-0.5 <$ [Fe/H] $< 0.64$) and subsolar
metallicities ([Fe/H] $\le -0.5$), respectively. Green-filled circles denote the five
newly identified Ba dwarfs in this work. The semicircular curves indicate
the boundaries separating thin-disk, thick-disk, and halo populations, with
$V_{\rm tot} = (U_{\rm LSR}^{2} + V_{\rm LSR}^{2} + W_{\rm LSR}^{2})^{1/2}$
less than 85~km\,s$^{-1}$ (blue curve) and 180~km\,s$^{-1}$ (red curve),
respectively \citep{Chen2004}.}
\label{fig:uvw}
\end{figure}

Figure~\ref{fig:uvw} presents the Toomre diagram, where the quantity
$(U_{\rm LSR}^{2}+W_{\rm LSR}^{2})^{1/2}$ is plotted against
$V_{\rm LSR}$. The semicircular curves mark the canonical
boundaries separating the thin- and thick disk populations,
corresponding to total velocities
$V_{\rm tot} = (U_{\rm LSR}^{2}+V_{\rm LSR}^{2}+W_{\rm LSR}^{2})^{1/2}$
of 85 and 180~km\,s$^{-1}$, respectively \citep{Chen2004}.
As shown in the figure, the majority of our stars belong to the Galactic
thin and thick disks, consistent with the findings of \citet{Gomez1997},
\citet{Pereira2011}, and \citet{Norfolk2019}. In contrast, 18 objects exhibit
kinematics characteristic of the Galactic halo. Furthermore, stars with
near-solar metallicities are found primarily in the thin and thick disks,
whereas those residing in the halo are predominantly metal-poor.
The five newly identified Ba dwarfs all have near-solar metallicities
and are located in the thin disk, in agreement with expectations from models
of Galactic chemical evolution \citep{Travaglio2001,Brusadin2013,Matteucci2021}.

\begin{figure}
\centering
\includegraphics[width=14cm]{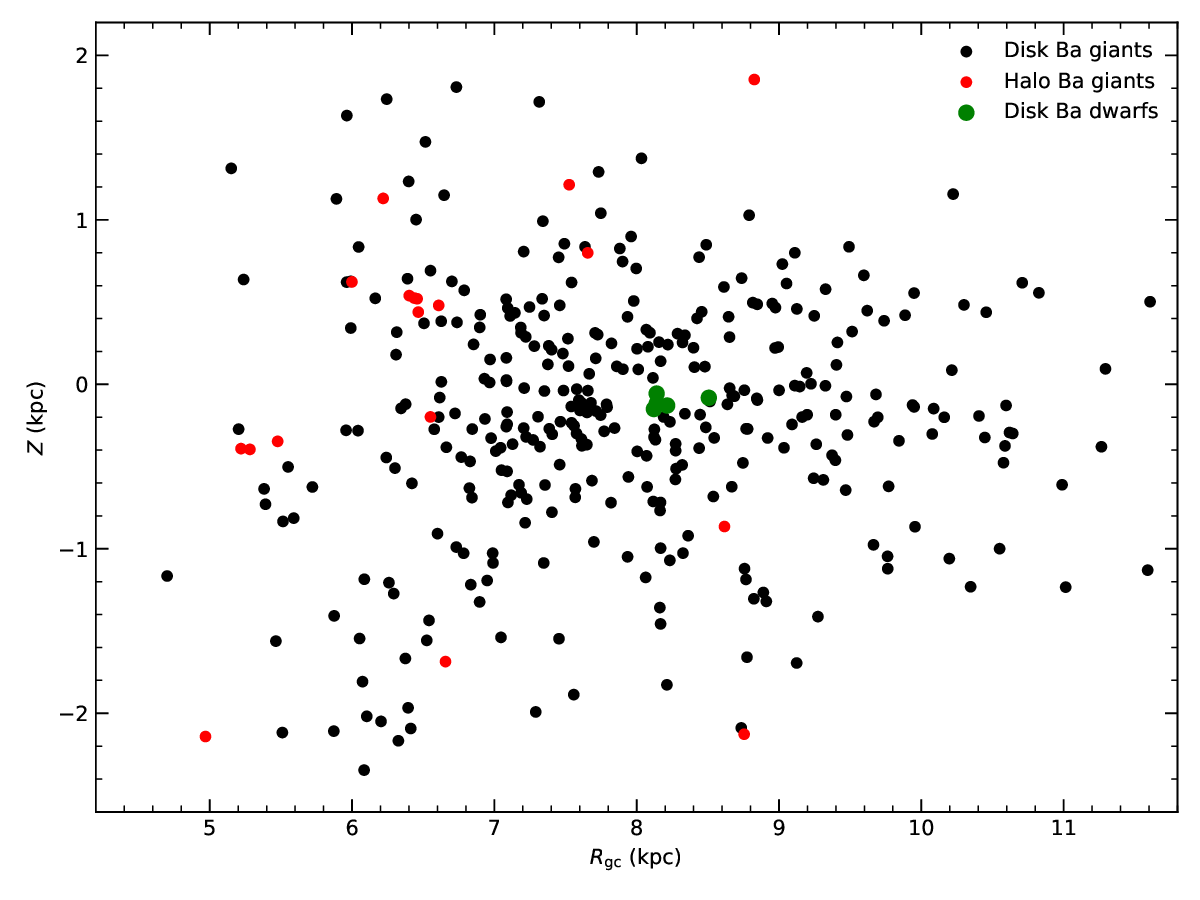}
\caption{Spatial distribution of the 367 Ba stars, mostly located
within 4~kpc, plotted as vertical height versus Galactocentric radius.
Filled black and red circles represent Ba giants associated with the
Galactic disk and halo, respectively. Filled green circles denote Ba dwarfs.}
\label{fig:rz}
\end{figure}

To examine the spatial distribution of the Ba stars, we plot their vertical
height ($Z$) as a function of Galactocentric radius ($R_{\rm gc}$) in
Figure~\ref{fig:rz}. The $X$, $Y$, and $Z$ Galactocentric coordinates are
taken from the GALAH DR4 catalog, with the origin defined at the Galactic center.
As expected from our distance cut ($d < 4$~kpc), the resulting 367 stars are
concentrated in the solar neighborhood in the $Z$-direction.
In contrast to the relatively scattered spatial distribution of the Ba
giants, the five Ba dwarfs exhibit a much more concentrated distribution and
are all located in the solar neighborhood. Interestingly, 18 stars classified
as halo members are located near the solar plane. This may be explained by their
current orbital phase, as halo stars can pass through the disk and temporarily
reside near $Z \approx 0$ during their Galactic orbits.

\subsection{Accreted versus in situ origin}

The stars in our Galactic halo preserve the record of its hierarchical assembly,
shaped by successive accretion events \citep{Gilmore2002,Majewski2017,Laporte2018}.
Distinguishing accreted stars from in situ stars is challenging, as debris from
ancient mergers is largely phase mixed. Conserved dynamical quantities, particularly
angular momentum and orbital energy, combined with stellar chemical
abundances, provide a powerful means to identify progenitor systems \citep{Naidu2020}.
Large-scale surveys have revealed both well-known structures, such as the Sagittarius
and Helmi streams, and recent discoveries like Gaia–Sausage–Enceladus (GSE),
demonstrating that past mergers have left discernible kinematic and chemical imprints
\citep{Belokurov2018,Helmi2018}. These chemodynamical signatures thus offer a
robust framework for tracing the origin of individual stars
\citep{Belokurov2020,Han2022,Dodd2023}.

\begin{figure}
\centering
\includegraphics[width=14cm]{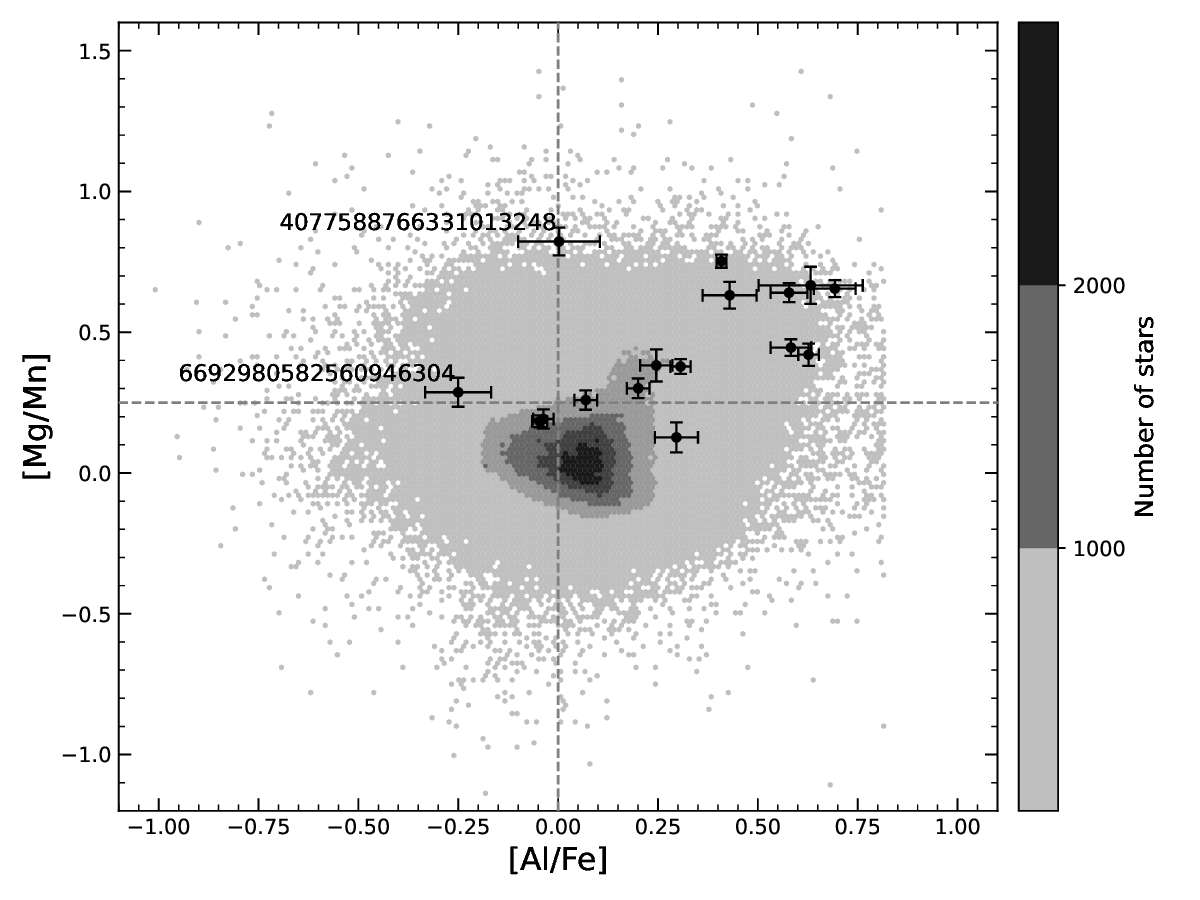}
\caption{[Mg/Mn] vs. [Al/Fe] for the halo Ba stars (black circles).
The vertical and horizontal dashed lines denote [Al/Fe] = 0 and [Mg/Mn] = 0.25,
respectively. For reference, all GALAH~DR4 stars are shown in the background as
filled gray circles.}
\label{fig:mgmn-alfe}
\end{figure}

The [Mg/Mn] versus [Al/Fe] plane provides a powerful diagnostic for distinguishing
accreted from in situ stellar populations. Mg predominantly traces enrichment from
core-collapse supernovae, Mn from Type Ia supernovae (SNe~Ia), and Al tends to be
under-abundant in stars originating from dwarf galaxies compared to in situ populations
\citep{Arcones2023}. Halo stars exhibiting [Mg/Mn] $>$ 0.25 and [Al/Fe] $<$ 0 are
typically linked to accreted systems \citep{Carrillo2024}. Based on the GALAH
abundance measurements, Figure~\ref{fig:mgmn-alfe} illustrates the [Mg/Mn]–[Al/Fe]
distribution of the halo Ba stars. Most of these stars are consistent with an in situ
origin, whereas two of them, 4077588766331013248 and 6692980582560946304,
are likely accreted from external systems when abundance uncertainties are taken
into account. This finding suggests that they may be remnants of past merger events.

To explore the possible substructure associations of the two likely accreted Ba stars,
we determined their locations in the three-dimensional integrals-of-motion space,
characterized by orbital energy ($E$) and the $z$-component of angular momentum ($L_{\rm z}$).
For comparison, the member-star catalog of different substructures was adopted from \citet{Dodd2023}.
To ensure consistency, we recalculated the $E$ and $L_{\rm z}$ values for the sample
stars of \citet{Dodd2023} using the same method applied to our targets. The astrometric
and kinematic parameters (e.g., R.A., decl., distance, proper motion, and radial velocity) for
both \citet{Dodd2023} and our sample stars were taken from Gaia~DR3 \citep{Gaia2023}.

\begin{figure}
\centering
\includegraphics[width=14cm]{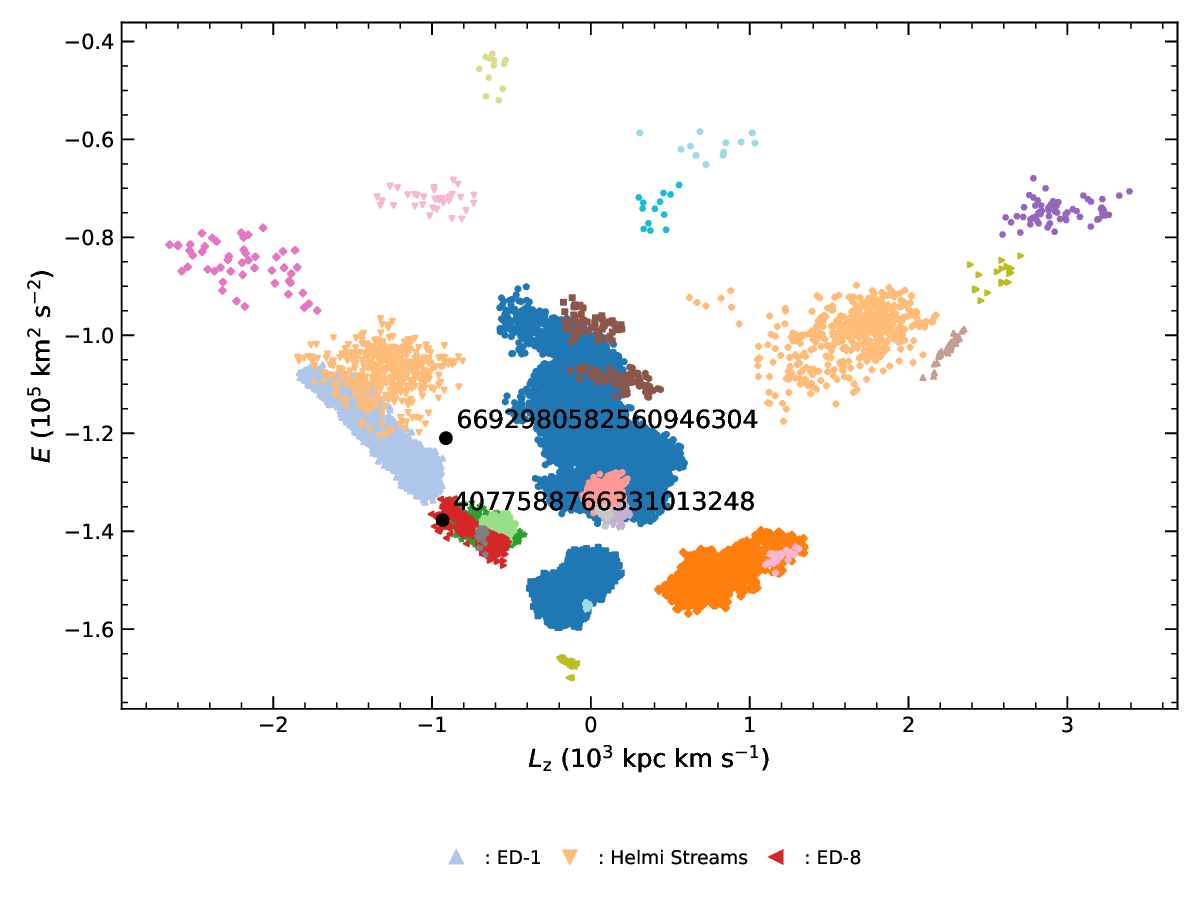}
\caption{$E$–$L_{\rm z}$ distribution of Galactic halo substructures. Black points
mark the positions of our two target stars. Colored points represent different
substructures, including Sagittarius, Aleph, GSE, Helmi streams, Thamnos, Sequoia,
Arjuna, I\'itoi, the in situ halo, EDs, and so on \citep{Naidu2020,Dodd2023}. The
substructures located near or overlapping our two sample stars are labeled,
illustrating the possible substructure membership of the two Ba stars.}
\label{fig:Lz_E}
\end{figure}

The calculations were performed in the Galactocentric reference frame implemented in
Astropy \citep{Astropy2013,Astropy2018,Astropy2022}. We adopted a solar position of
$R_\odot = 8.21$~kpc \citep{McMillan2017,Bennett2019} and a solar velocity relative to
the LSR of $(U, V, W)_{\odot} = (11.1, 12.24, 7.25)$~km~s$^{-1}$ \citep{Schonrich2010},
with $V_{\rm LSR} = 233.1$~km~s$^{-1}$ \citep{McMillan2017}. The coordinate system is
right-handed, such that prograde and retrograde orbits correspond to $L_{\rm z} < 0$
and $L_{\rm z} > 0$, respectively. Orbital quantities, including actions,
eccentricities, and energies, were computed using Gala with its default {\tt MilkyWayPotential}
\citep{Price2017}. The resulting distributions are shown in Figure~\ref{fig:Lz_E}.
From the figure, we can see that the Ba star 4077588766331013248 lies within the region of
substructure ED-8 identified by \citet{Dodd2023}. This provides additional evidence that
this star may have an accreted origin, consistent with the chemical abundance signature
shown in Figure~\ref{fig:mgmn-alfe}. For the sample star 6692980582560946304, its location is
close to the Helmi streams and ED-1 substructures \citep{Dodd2023}, or possibly the
high-$\alpha$ disk \citep{Naidu2020}. In this case, the analysis based on
Figures~\ref{fig:mgmn-alfe} and \ref{fig:Lz_E} makes it difficult to determine whether
the star has an in situ or accreted origin.

\subsection{Abundances of $s$-process elements and carbon}

\begin{figure}
\centering
\includegraphics[width=14cm]{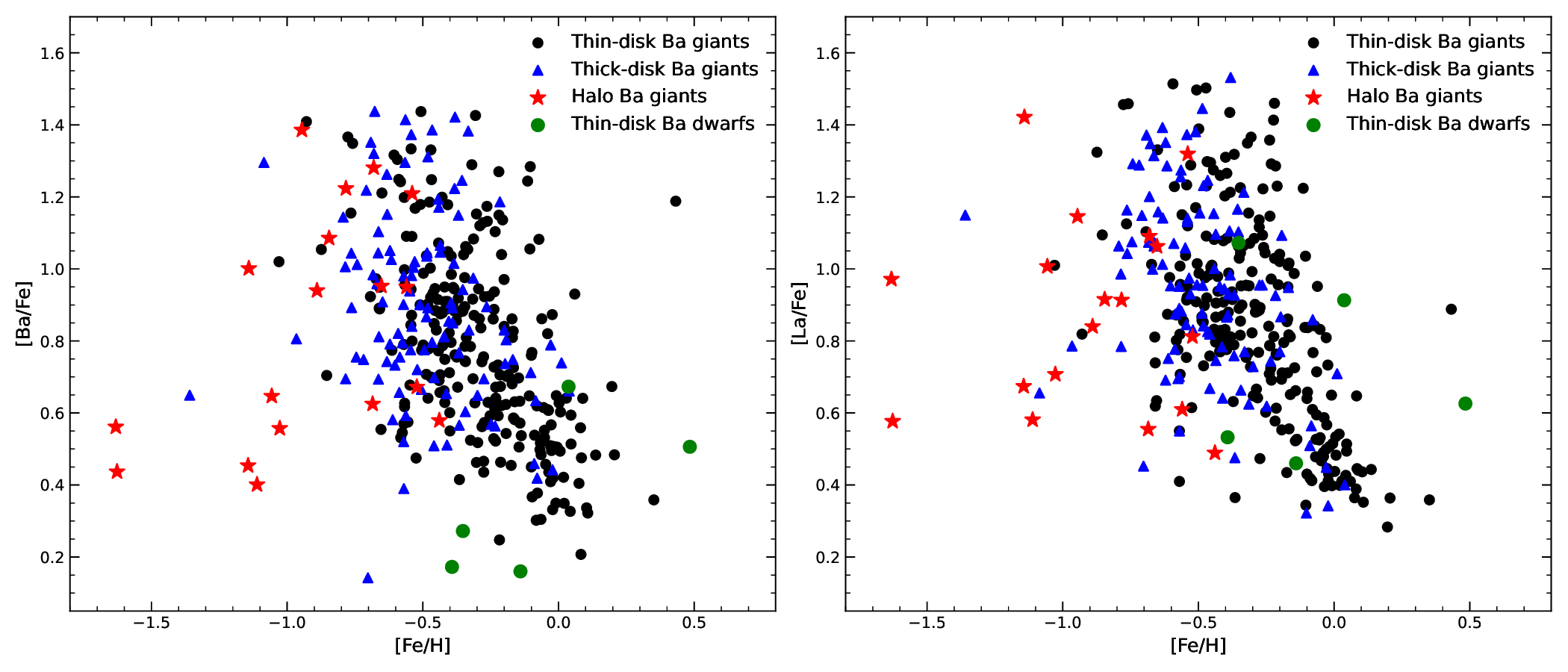}
\caption{Abundance ratios [i/Fe] as a function of [Fe/H] for the two representative
$s$-process elements, Ba and La, in our Ba stars. Filled black circles, blue triangles, and
red stars represent Ba giants belonging to the thin disk, thick disk, and halo, respectively.
Filled green circles denote Ba dwarfs.}
\label{fig:sfe}
\end{figure}

The chemical abundances of $s$-process elements serve as important tools in studying
nucleosynthesis processes and the formation history of Ba stars. Figure~\ref{fig:sfe}
shows the [i/Fe] ratios of two representative $s$-process elements, Ba and La, for
the 486 stars in our sample. Most Ba stars exhibit varying degrees of $s$-process
overabundance. At [Fe/H] $\gtrsim -1.0$, the majority of stars are associated with
the Galactic thin and thick disks. As metallicity increases, the [i/Fe] ratios
gradually decline, indicating that $s$-process and iron-peak (e.g., Cr, Mn, Fe,
Co, and Ni) elements are not produced in the same astrophysical sites. The $s$-process
elements are primarily synthesized in low- and intermediate-mass AGB stars, whereas
iron-peak elements are predominantly produced by nuclear fusion in massive stars and SNe~Ia
\citep{Sneden2008,Arcones2023}. At lower metallicities ([Fe/H] $\lesssim -1.0$),
the [i/Fe] ratios drop sharply, likely reflecting the limited contribution of
$s$-process nucleosynthesis in the early Galaxy \citep{Travaglio1999,Burris2000}.
Furthermore, as shown in the figure, the [i/Fe] ratios of Ba dwarfs show
systematic differences compared to the main distribution of Ba giants at similar
metallicities. At [Fe/H] $\lesssim 0$, Ba dwarfs tend to exhibit lower [i/Fe] ratios
than Ba giants, whereas at [Fe/H] $\gtrsim 0$, the trend appears to be reversed.
This behavior may suggest that Ba dwarfs and Ba giants have experienced different
chemical enrichment histories.

\begin{figure}
\centering
\includegraphics[width=14cm]{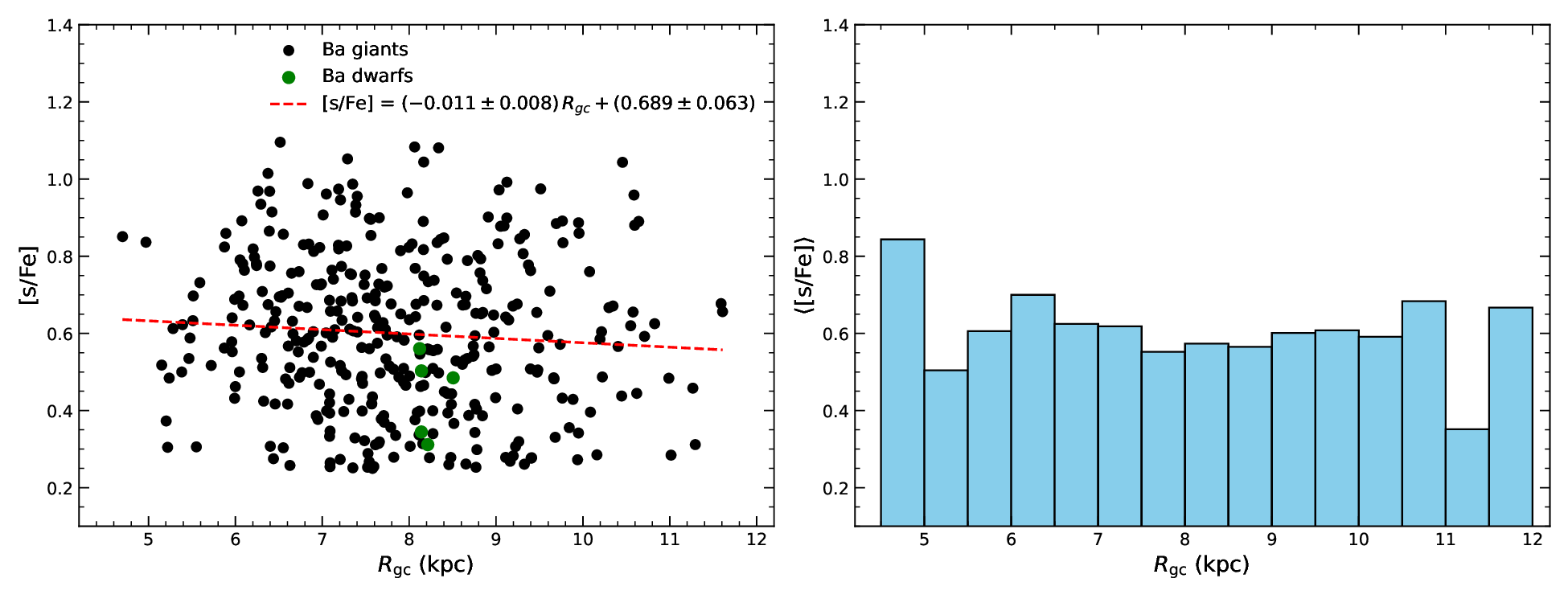}
\caption{
Abundance ratios [s/Fe] of the Ba stars as a function of Galactocentric
radius $R_{\rm gc}$. Left panel: individual [s/Fe] ratios of Ba giants and dwarfs
are shown as filled black and green circles, respectively, with the red line indicating
the linear fit to the entire sample. Right panel: mean [s/Fe] values averaged in
bins of 0.5~kpc in $R_{\rm gc}$.}
\label{fig:rgc-sfe}
\end{figure}

To examine whether there is a radial dependence in the efficiency of $s$-process
enrichment among the Ba stars, we investigated the behavior of the [s/Fe] ratios as a
function of Galactocentric radius $R_{\rm gc}$. Specifically, we plotted [s/Fe] versus
$R_{\rm gc}$ and computed the mean [s/Fe] values in bins of 0.5~kpc in $R_{\rm gc}$, as shown in
Figure~\ref{fig:rgc-sfe}. As illustrated in the left panel, the [s/Fe] ratios display
a nearly flat trend with increasing $R_{\rm gc}$, with a fitted slope of $-0.011 \pm 0.008$
and a large scatter at all radii. The right panel further confirms this result, showing
that the mean [s/Fe] values in different $R_{\rm gc}$ bins do not exhibit any significant
increasing or decreasing trend. Overall, we find no clear evidence for a radial dependence
of the $s$-process enrichment efficiency within the spatial range probed by our sample.
For the five Ba dwarfs, their $s$-process enrichment levels are generally lower
than the mean value of the Ba giants, and they are all located in the solar
neighborhood in terms of Galactocentric radius. Owing to the small number
of Ba dwarfs, no firm conclusion can be drawn regarding their radial behavior.

\begin{figure}
\centering
\includegraphics[width=14cm]{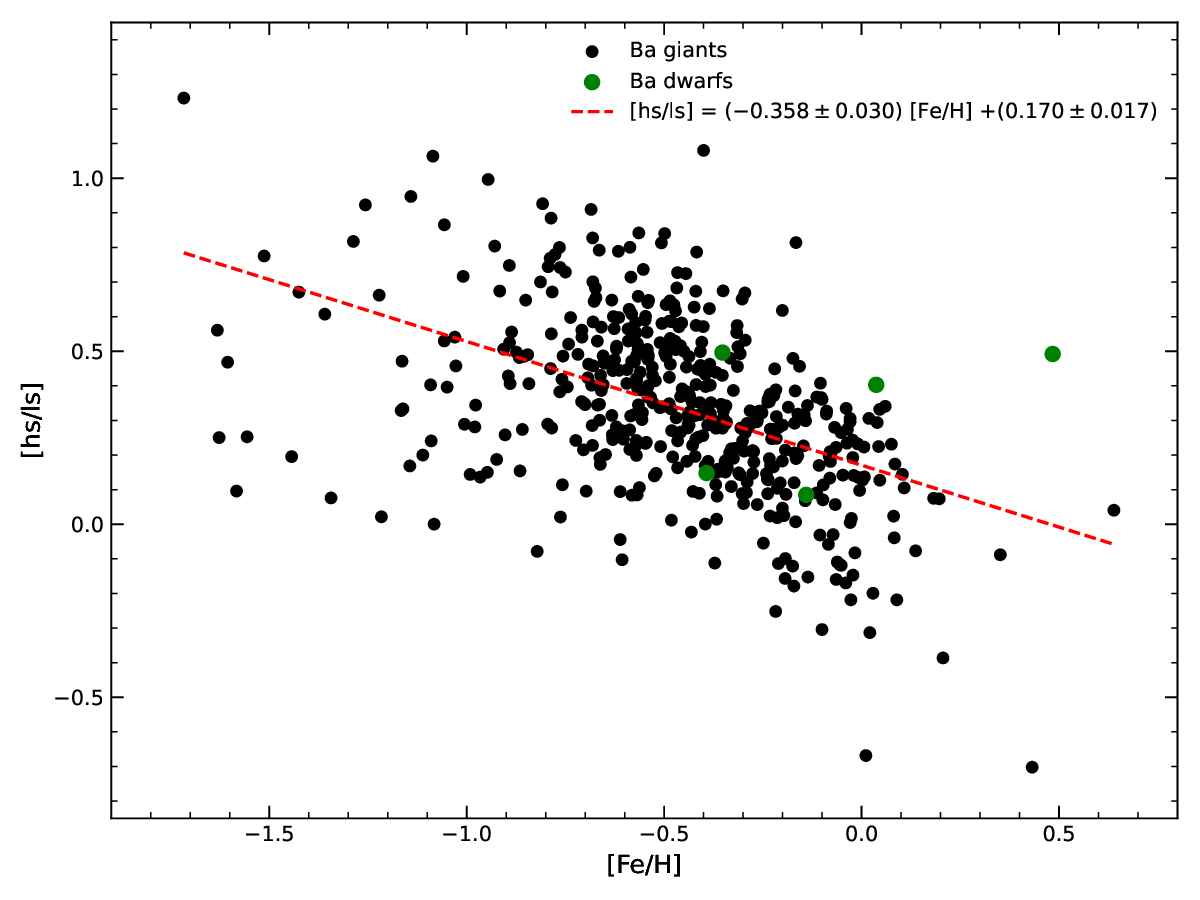}
\caption{Abundance ratio [hs/ls] as a function of [Fe/H] for the Ba star sample.
The dashed line represents the best-fit linear regression to the data.}
\label{fig:hsls}
\end{figure}

The [hs/ls] ratio, constructed from heavy $s$-process (Ba, La, Ce, Nd) and
light $s$-process (Sr, Y, Zr) elements, serves as a sensitive probe of $s$-process
efficiency and places strong constraints on AGB nucleosynthesis models.
Figure~\ref{fig:hsls} shows [hs/ls] as a function of [Fe/H] for our sample stars.
A clear decreasing trend with metallicity is seen, in agreement with previous
studies \citep[e.g.,][]{Cseh2018,Roriz2021b,Yang2024}. A linear fit to
the data yields $[\rm hs/ls] = (-0.358 \pm 0.030)\,[\rm Fe/H] + (0.170 \pm 0.017)$.
This trend indicates that heavy $s$-process elements dominate the neutron-capture
yields at low metallicities ([Fe/H] $< -0.6$), while light $s$-process elements
become more prevalent at higher metallicities \citep{Busso2001}. Negative [hs/ls]
values reflect a relative enhancement of light $s$-process elements at the
corresponding metallicities.
The [hs/ls] distributions of Ba dwarfs and Ba giants do not show any obvious
systematic differences. However, the present Ba dwarf subsample does not yet allow
a statistically robust assessment of potential differences in their [hs/ls] properties.

Furthermore, the [hs/ls] scatter decreases with increasing metallicity,
which can be interpreted as a consequence of the gradual homogenization of
the interstellar medium over time, driven by the mixing of chemical yields
from multiple nucleosynthetic sites, including the main $s$-process in AGB
stars and the weak $s$-process in massive stars \citep{Travaglio1999,Travaglio2004}.

\begin{figure}
\centering
\includegraphics[width=14cm]{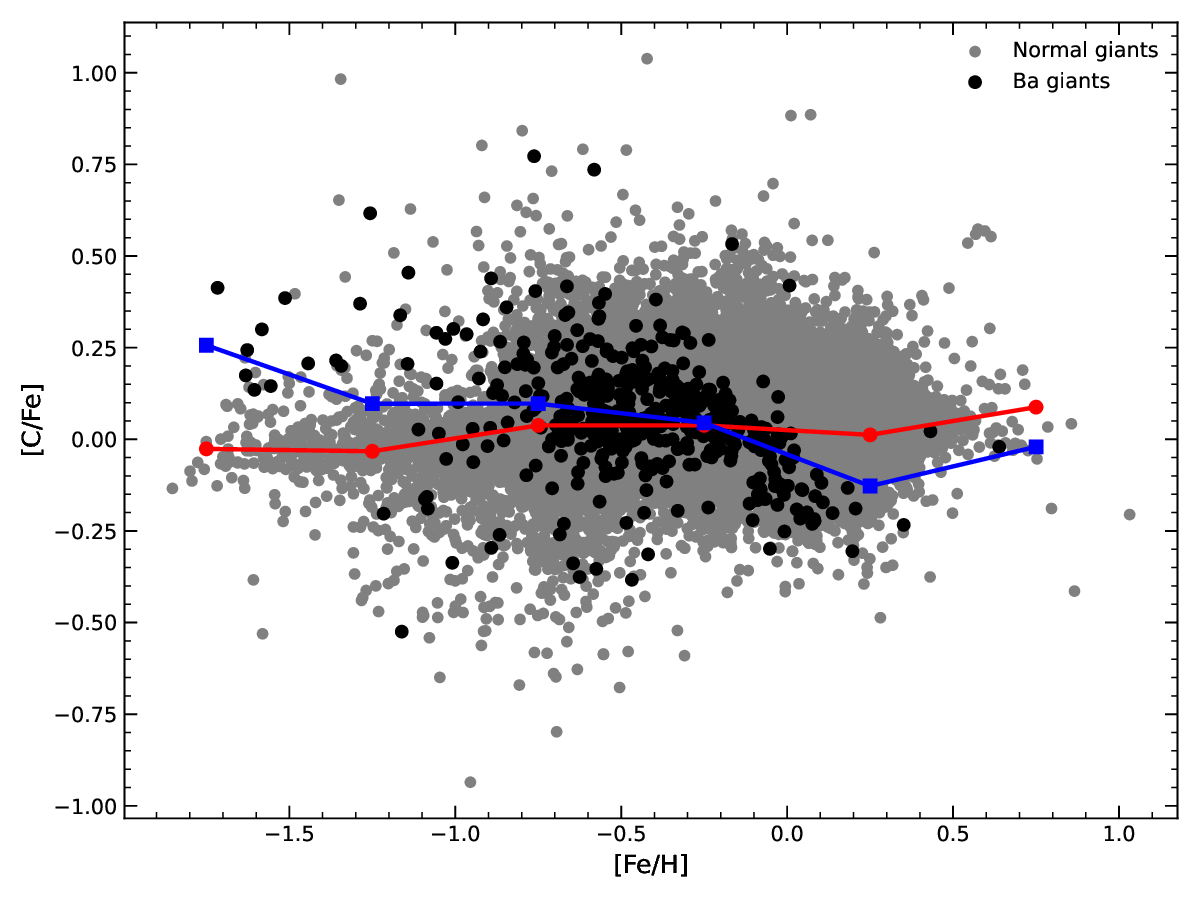}
\caption{Abundance ratio [C/Fe] as a function of [Fe/H] for the Ba giants (black circles)
and normal giants (filled gray circles). The blue and red curves refer to the average [C/Fe]
ratios in metallicity bins of 0.50 for Ba giants and normal giants, respectively.}
\label{fig:cfe}
\end{figure}

In addition to the $s$-process elements, the abundance behavior of the light
element carbon (C) is of particular interest in Ba stars. As a tracer that is highly
sensitive to mixing processes during different stages of stellar evolution, carbon
provides valuable insight into the nature of the former companion that polluted the
Ba stars during its thermally pulsing (TP-) AGB phase \citep{Karinkuzhi2021b,Roriz2025}.
In this work, because reliable carbon abundances
are not available for the five Ba dwarfs in GALAH~DR4, we focus our analysis on the
Ba giants. Using homogeneous data from GALAH~DR4, we compare the carbon abundances of
Ba giants with those of normal giants. As shown in Figure~\ref{fig:cfe}, the blue
and red curves represent the mean [C/Fe] values in metallicity bins of 0.50 for
the Ba giants and normal giants, respectively. The [C/Fe] ratios of Ba giants exhibit
an overall decreasing trend with increasing [Fe/H], in agreement with previous studies
\citep{Kong2018,Roriz2024}. In contrast, the [C/Fe] ratios of normal giants show a
weakly increasing or nearly flat trend at [Fe/H] $\gtrsim -1.5$, consistent with the
results of \citet{Horta2023}. The weaker increase in [C/Fe] at [Fe/H] $>-1.5$, compared
to lower metallicities, may reflect the increasing contribution of SNe~Ia to iron-peak
elements without a corresponding production of carbon.

When comparing the mean [C/Fe] values of Ba giants and normal giants, we find that Ba
giants display systematically higher [C/Fe] ratios at metallicities below
[Fe/H] $\approx -0.3$, consistent with carbon enrichment through mass transfer from an
AGB companion. At higher metallicities, however, the opposite behavior is observed,
with Ba giants exhibiting lower [C/Fe] ratios than normal giants. This metallicity-dependent
pattern suggests that carbon enhancement associated with $s$-process enrichment is more
pronounced at low metallicities, while at higher metallicities it may be diluted by a
combination of stellar evolutionary mixing and the increased iron enrichment from SNe~Ia.
Over the full metallicity range, the average [C/Fe] of the Ba giants is higher than that
of the normal giants by approximately $\sim$0.04, which is significantly smaller than
the enhancement reported by \citet{Roriz2025}.

\subsection{Binarity}

In our sample, the Ba giants and dwarfs are not sufficiently evolved to undergo
internal $s$-process nucleosynthesis, and are therefore considered products of
binary (or possibly triple) interactions. Ba stars are thought to form when
a former AGB companion--—now a WD--—enriches the primary star with
$s$-process material through mass transfer. Using spectral residual analysis,
\citet{Bunder2024} searched for binary signatures among GALAH DR4 targets and
flagged likely binary systems by reporting the radial velocity of the primary
source ($rv\_comp\_1$) and of the potential secondary source ($rv\_comp\_2$)
in the catalog. Among our 486 Ba stars, only 74 have measured $rv\_comp\_2$ values.
To further investigate binarity, we compared the radial velocities from
GALAH DR4 with those from Gaia DR3. Stars with velocity residuals exceeding
the combined observational uncertainties were classified as showing significant
radial-velocity variability. Unfortunately, only 271 stars in our sample
exhibit such variations.

Although radial-velocity monitoring is an important diagnostic for confirming
the binarity of Ba stars, it can be challenging to detect. This difficulty may
arise from long orbital periods \citep{Escorza2019,Escorza2020,Jorissen2019},
nearly pole-on configurations, or highly eccentric orbits where significant
velocity changes occur only over a small fraction of the orbital phase
\citep{Pourbaix2004,deCastro2016}. Ultraviolet (UV) photometric excess in the
spectral energy distribution (SED) of a star can serve as an indicator of a WD
companion \citep{Vitense2000,Shetye2020}. We constructed SEDs for our Ba stars
using the Virtual Observatory SED Analyser (VOSA; \citealt{Bayo2008}) to search
for near-UV (NUV; centered at 2271~\AA) excess and thereby diagnose the presence
of WD companions.

Prior to this analysis, we employed the IDL-based code KMOD, which is widely used
in stellar abundance studies \citep[e.g.,][]{Meszaros2013,Yong2013}, to interpolate
the ATLAS9 model atmosphere grid \citep{Castelli2004}. This interpolation provided
the atmospheric model corresponding to the target stellar parameters. Using these
models, we generated synthetic spectra with SPECTRUM \citep{Gray1994}. For the
input chemical composition, the abundances of neutron-capture elements were adopted
from GALAH DR4, as they often differ significantly from solar values, whereas
abundances of lighter elements (up to the iron-peak) were kept at solar values
\citep{Asplund2009}.

\begin{figure}
\centering
\includegraphics[width=14cm]{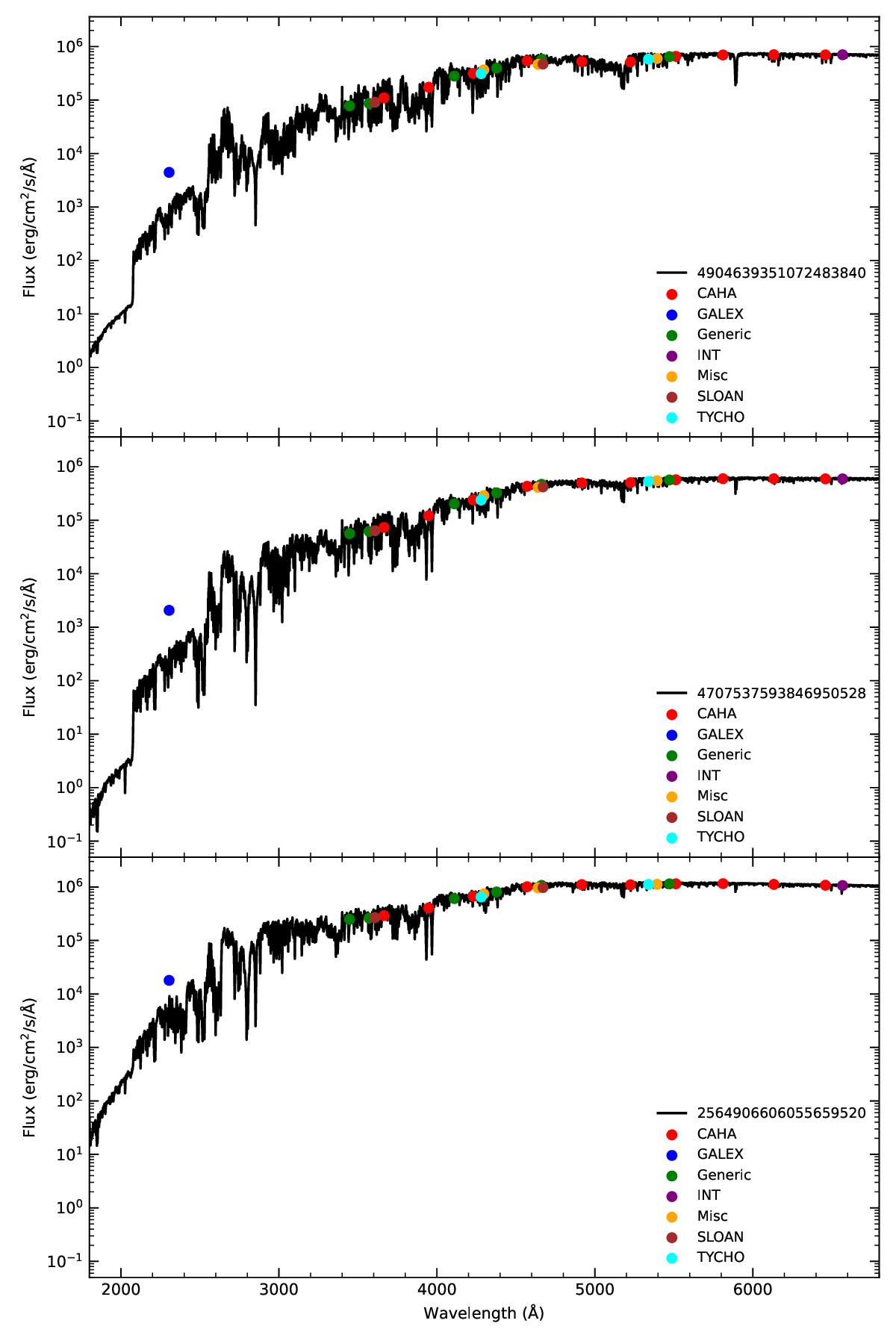}
\caption{SEDs of one Ba dwarf (4904639351072483840) and two Ba giants (4707537593846950528
and 2564906606055659520), shown as illustrative examples and constructed using VOSA.
The NUV photometric data are from GALEX, shown as filled blue circles in the figure.
For comparison, photometric measurements in other bands, including those from CAHA,
SDSS, and additional surveys, are also plotted. The model spectra are taken from
the ATLAS9 grid of stellar atmospheres.}
\label{fig:vosa}
\end{figure}

Using these synthetic templates, we evaluated the NUV fluxes of all 486 Ba stars and
found significant NUV excesses in all five Ba dwarfs and in nearly the entire
Ba giant sample. Figure~\ref{fig:vosa} presents three representative examples (one
Ba dwarf, 4904639351072483840, and two Ba giants, 4707537593846950528 and
2564906606055659520), showing the GALEX NUV measurements \citep{Bianchi2017} together
with photometry in additional bands (e.g., $U$, $B$, $V$, $u$, $v$, $g$, $b$, and $y$)
from surveys such as CAHA, Generic, Sloan Digital Sky Survey (SDSS), and others for
comparison. The optical photometry at wavelengths longer than 3400~\AA\ closely follows
the model spectrum, indicating no flux excess in these bands. In contrast, the observed
GALEX NUV flux is clearly higher than predicted by the model, providing further evidence
that these stars host WD companions, consistent with the widely accepted formation
scenario for Ba stars. As shown in the four panels of the figure, the NUV excesses
observed in the SEDs of Ba dwarfs exhibit characteristics similar to those seen in Ba giants.

Six stars, however, show no obvious NUV excess. Among them, 4657277066293133056,
3317356770048876672, 6090437043630921472, 4285878256152184832, and 4199611978282443264
can still be classified as binaries based on the presence of the $rv\_comp\_2$ flag or
on detectable radial-velocity variations between GALAH~DR4 and Gaia~DR3
\cite[see Table~B1 in][]{Bunder2024}. For the Ba giant 6053735173729807872, no
convincing evidence of binarity is found. Its orbit or geometric configuration may be
unfavorable for detection, or the effective temperature of a potential WD
companion may be too low to produce a measurable NUV excess. If the star is
truly single, the origin of its Ba and La overabundances requires further investigation.

\subsection{Companion mass estimates}

Based on nucleosynthesis models of low-mass stars, \citet{Cristallo2011} investigated
the $s$-process in AGB stars and developed the FRUITY database. Their models cover
stellar masses of 1.3-–6.0~$M_{\odot}$ and metallicities from $Z = 10^{-5}$ to $3 \times 10^{-2}$.
In these models, the main neutron source is assumed to be the radiative burning of $^{13}$C,
and the efficiency of the third dredge-up (TDU) decreases with increasing metallicity.
For a given metallicity, the surface enrichment is determined primarily by the amount of
dredged-up material and the depth of the convective envelope.

In this work, we inferred the masses of the Ba star companions by comparing the observed
neutron-capture element abundances of our sample with the predictions from the FRUITY AGB
nucleosynthesis models. Direct comparison with model yields is not straightforward, as the
transferred material from the former AGB companion becomes diluted in the envelope of the
Ba star. This dilution is driven by both the mass-transfer process itself and subsequent
internal mixing, primarily during the first dredge-up. The degree of dilution can be
expressed as $\mathrm{dil} = (M_{\rm AGB,trans} + M_{\rm Ba,env})/M_{\rm AGB,trans}$
\citep{Bisterzo2012}, where $M_{\rm AGB,trans}$ is the mass accreted from the companion
and $M_{\rm Ba,env}$ is the mass of the Ba star's envelope.

The abundance patterns of AGB models were computed for each Ba star using the
approach described by \citet{Cseh2022}. For a given element, the diluted abundance,
[X/Fe]$_{\rm dil}$, was obtained using

\begin{equation}\label{dilution}
\rm [X/Fe]_{dil} = \log \left(10^{[X/Fe]_{\rm ini}} (1-\delta) + 10^{[X/Fe]_{\rm AGB}} \delta \right),
\end{equation}

where [X/Fe]$_{\rm ini}$ denotes the initial abundance ratio, and [X/Fe]$_{\rm AGB}$
corresponds to the post-AGB yields predicted by the models. The dilution factor
$\delta$ (=1/dil) was determined by scaling the model [Ba/Fe] ratio to match that
of the observed star, which is different from \citet{Cseh2022}, who derived
the $\delta$ values by scaling the model [Ce/Fe] ratio to the observed [Ce/Fe].
This factor was then applied to compute [X/Fe]$_{\rm dil}$ for all other elements. 
To identify the most representative model for each Ba star,
we first restricted the comparison to AGB models with metallicities close to that
of the star, and subsequently minimized the $\chi^{2}$ difference between the
observed and model-predicted abundances. Solutions with $\delta < 1$ were retained,
since $\delta = 1$ would correspond to a physically implausible case of pure,
unmixed AGB ejecta.

\begin{figure}
\centering
\includegraphics[width=14cm]{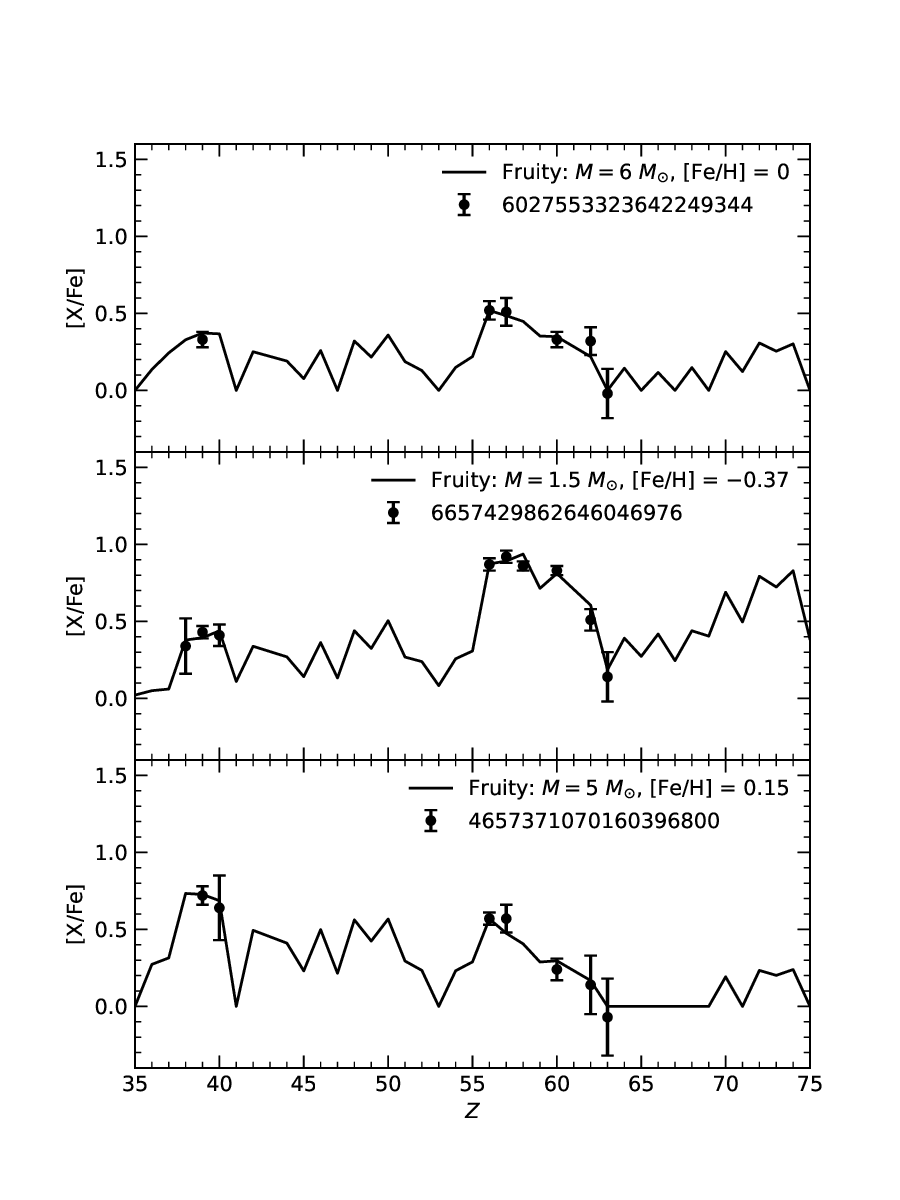}
\caption{Comparison between the observed abundance patterns of the three sample
stars (filled circles with error bars) and the best-fitting FRUITY AGB model
predictions (curves) after applying dilution.}
\label{fig:fruity}
\end{figure}

\begin{figure}
\centering
\includegraphics[width=14cm]{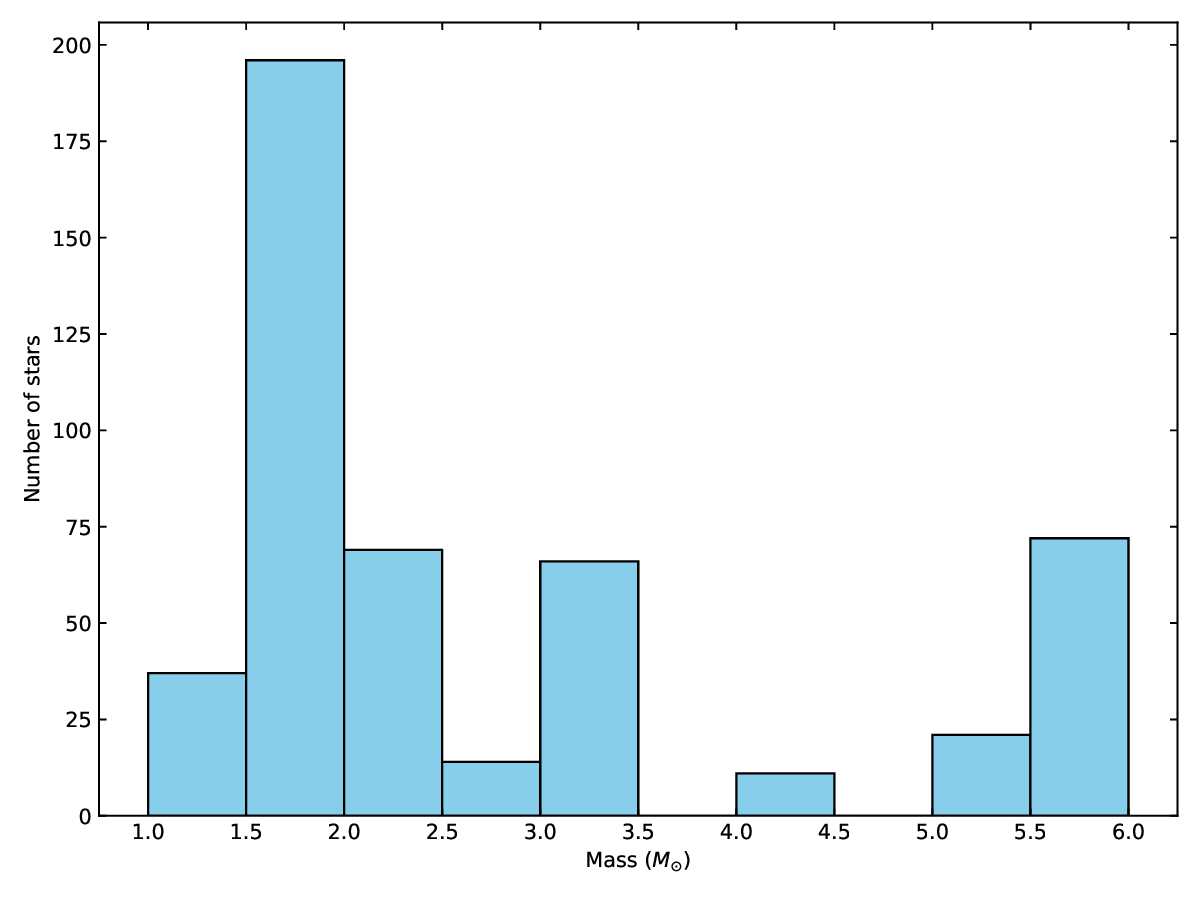}
\caption{Histogram illustrating the mass distribution of companions to Ba stars.}
\label{fig:mass-num}
\end{figure}

As illustrative cases, Figure~\ref{fig:fruity} presents the abundance pattern
comparisons between the three selected Ba stars and their corresponding best-fit
AGB models. The diluted yields from AGB stars of the inferred masses reproduce
the observed elemental distributions remarkably well. The inferred companion masses
for the entire sample are summarized in Table~\ref{tab:catlog} (column~24).
Their distribution across different mass intervals is shown in Figure~\ref{fig:mass-num},
which highlights that the majority of Ba star companions are low-mass
($M = 1.5$–-$3.5~M_{\odot}$) TP-AGB stars. These findings align well with previous
studies \citep[e.g.,][]{Allen2006a,Cseh2018,Goswami2023,Roriz2024}, reinforcing the view
that such low-mass AGB stars are the dominant polluters in Ba star systems.

Nevertheless, as shown in Table~\ref{tab:catlog} and Figure~\ref{fig:mass-num},
there are 93 Ba stars---comprising two Ba dwarfs and 91 Ba giants--—whose inferred companion
masses exceed $5.0~M_{\odot}$, despite the majority of the sample being consistent with
pollution by low-mass AGB companions. This result suggests that intermediate-mass stars
evolving through the AGB phase and producing $s$-process elements may also contribute to
the chemical enrichment of Ba binaries, in agreement with previous studies
\citep[e.g.,][]{Jorissen2019,Roriz2021a,Cseh2022}. In addition, we note that a small
number of systems exhibit inferred AGB companion masses that are smaller than the present
masses of the Ba stars  ($M_{\rm AGB} < M_{\rm Ba}$). The companion masses estimated in
this work correspond to the initial masses of the former AGB stars, inferred by fitting
the observed $s$-process abundance patterns with the FRUITY AGB nucleosynthesis models,
rather than to their present-day remnant masses. During binary evolution, mass transfer
through Roche-lobe overflow and/or wind mass accretion can substantially increase the
mass of the initially less massive star. As a consequence, a mass-ratio reversal may
occur, such that the initially more massive star (the AGB progenitor) becomes less massive
after losing most of its envelope, while the initially less massive star (now observed as
the Ba star) gains mass and becomes the more massive component. Recent three-dimensional
hydrodynamical simulations of Roche-lobe overflow by \citet{Ryu2025} further suggest that
the present-day mass ordering of binary components may not necessarily reflect their initial
masses, particularly in systems that have undergone substantial mass exchange. Consequently,
cases in which the inferred AGB progenitor mass is comparable to or even lower than the
current mass of the Ba star may plausibly reflect the evolutionary history of mass transfer
in these systems.


\section{Conclusions}

In this work, we identified a sample of 486 Ba stars from the GALAH~DR4 dataset
using high-resolution spectra and precise abundance measurements. The sample was
constructed based on the enhancement criteria involving typical $s$-process elements
(Ba and La) relative to the $r$-process element (Eu), as proposed by \citet{Beers2005},
and the [s/Fe] criterion by \citet{deCastro2016}. To further ensure the purity of
the Ba star sample, we applied the signed distance method proposed by
\citet{Karinkuzhi2021a} to remove possible non-Ba contaminants. Based on the above
methodology, we obtained the largest Ba star sample to date. By presenting the
distribution of $T_{\rm eff}$ versus $\log g$, we identified five new Ba dwarfs
(4904639351072483840, 4644665049364988928,
3220808727029495680, 5480510146667329152, and 6810035273351899648),
significantly expanding the previously small known population of such objects.
Based on this large Ba star sample, we analyzed their kinematic properties, progenitor
systems, $s$-process abundance characteristics, and binarity. Our main conclusions
are as follows:

\begin{enumerate}
\item Using reliable parallaxes, proper motions, and radial velocities from Gaia~DR3
and StarHorse, we derived the three Galactic velocity components ($U$, $V$, $W$) for 367
Ba stars. Most of these stars belong to the thin or thick disk populations, while a
small subset (18 stars) exhibits halo-like kinematics. Stars with near-solar metallicities
are mainly distributed within the Galactic disk, whereas the halo members are predominantly
metal-poor. The five newly identified Ba dwarfs are characterized by near-solar
metallicities and thin-disk kinematics, consistent with Galactic chemical evolution models.
The spatial distribution of the 367 Ba stars further shows that the 18 halo objects are
located near the solar neighborhood, likely because they are currently passing through
the disk near the sun along their Galactic orbits.

\item To investigate whether the halo Ba stars originated from accreted systems or
formed in situ, we analyzed their chemical abundances. Using the criteria of [Mg/Mn]
$>$ 0.25 and [Al/Fe] $<$ 0, which are typically associated with accreted populations
\citep{Carrillo2024}, we found that most Ba stars in our sample are likely of in situ
origin. However, considering abundance uncertainties, two stars—--4077588766331013248
and 6692980582560946304--—appear to be consistent with accreted populations. By combining our
data with the substructure catalog of \citet{Dodd2023}, we further calculated the
orbital angular momentum and total energy for these two stars. The star 4077588766331013248
is located within the region of substructure ED-8 identified by \citet{Dodd2023},
supporting an external (accreted) origin. In contrast, for 6692980582560946304, its
position in the $E$–$L_{\rm z}$ diagram makes it difficult to clearly distinguish
between an in situ and an accreted origin.

\item Enrichment in the $s$-process elements Ba and La is ubiquitous among the samples,
though to varying degrees. The decline of [Ba/Fe] and [La/Fe] ratios toward higher
metallicities ([Fe/H] $\gtrsim -1.0$) suggests that the $s$-process elements originate
from nucleosynthetic sites distinct from those of the iron-peak elements. The sharp
decrease in [i/Fe] ratios at low metallicities likely reflects the limited
contribution of $s$-process nucleosynthesis in the early Galaxy. At comparable
metallicities, the [i/Fe] ratios of Ba dwarfs tend to deviate from the locus defined
by Ba giants. No statistically significant radial dependence of the $s$-process enrichment
efficiency is found within the spatial extent of our sample. Furthermore, the observed
decline in the [hs/ls] ratio with increasing metallicity suggests enhanced neutron-capture
efficiency at lower metallicities. The greater dispersion of [hs/ls] at low [Fe/H]
relative to high [Fe/H] likely reflects the progressive homogenization and mixing of the
interstellar medium throughout Galactic evolution. No significant systematic
differences are observed between the [hs/ls] distributions of Ba dwarfs and Ba giants.
In addition, the [C/Fe] ratios of Ba giants show an overall decreasing trend with
increasing [Fe/H].

\item Observations to date show that Ba stars include both giants and dwarfs, none of
which are sufficiently evolved to synthesize $s$-process elements internally.
Radial-velocity variations and NUV excesses revealed by SED analyses confirm their binary
nature, indicating the presence of white-dwarf companions across the sample, with the
exception of the Ba giant 6053735173729807872. For this star, no clear signatures
of binarity are detected. If it is indeed part of a binary system, the lack of observable
binary characteristics may be due to an orbital configuration unfavorable for detection or
to a relatively cool companion incapable of producing a detectable NUV excess. Conversely,
if the star is truly single, the origin of its $s$-process enhancements requires further
investigation. The NUV excesses detected in the SEDs of Ba dwarfs are comparable to
those observed in Ba giants, suggesting a similar underlying origin.

\item For the 485 confirmed Ba binaries, companion masses inferred from FRUITY AGB models
predominantly fall in the range of 1.5–-3.5~$M_{\odot}$, consistent with low-mass AGB stars
that have undergone TDU episodes. Nevertheless, a number of Ba stars,
comprising both dwarfs and giants, have inferred companion masses above 5.0~$M_{\odot}$.
This indicates that intermediate-mass stars undergoing the AGB phase and synthesizing
$s$-process elements can also play a role in enriching Ba binaries. The dilution-corrected
AGB model yields successfully reproduce the observed abundance patterns, providing strong
support for the mass-transfer scenario for the formation of Ba stars.
\end{enumerate}

We hope that this catalog will provide a valuable foundation for future studies of
large Ba star samples, including binary-orbit monitoring and chemodynamical modeling.
Expanding the sample size will be crucial for gaining deeper insights into $s$-process
nucleosynthesis in AGB stars and the mass-transfer mechanisms operating in binary systems.

\section*{Data Availability}

This paper made use of the publicly available GALAH DR4 (https://www.galah-survey.org),
Gaia DR3 data (https://www.cosmos.esa.int/gaia), and StarHorse catalogs \citep{Queiroz2023}.
The data underlying this paper are available in machine-readable form.

\section*{Code Availability}
We use the standard data analysis tools as follows: TOPCAT \citep{Taylor2005},
Matplotlib \citep{Hunter2007}, NumPy \citep{Harris2020}, SciPy \citep{Virtanen2020},
Jupyter \citep{Kluyver2016}, Astropy \citep{Astropy2013,Astropy2018,Astropy2022},
and Gala \citep{Price2017}.

\begin{acknowledgments}
We thank the referee for the constructive comments and suggestions, which have
significantly improved the manuscript. This work is supported by the National
Natural Science Foundation of China (NSFC) under grant Nos. 12588202, 12273055,
and 12503024; the National Key R\&D Program of China under grant Nos. 2023YFE0107800
and 2024YFA1611902; the China Manned Space Program under grant No. CMS-CSST-2025-A13;
the Ministry of Science and Technology of China under grant No. 2023FY101104;
the International Partnership Program of the Chinese Academy of Sciences under grant
No. 178GJHZ2022040GC; the Independent Research Program of the National
Astronomical Observatories, Chinese Academy of Sciences under grant
No. E4ZB0301; the Strategic Priority Research Program of the Chinese Academy of
Sciences under grant No. XDB1160301; and the Fundamental Research Funds of China
West Normal University under grant No. 22kE041. We acknowledge the support from
the 2~m Chinese Space Station Telescope project.
This work made use of the Fourth Data Release of the GALAH Survey \citep{Bunder2024}.
The GALAH Survey is based on data acquired through the Australian Astronomical
Observatory, under programs: A/2013B/13 (The GALAH pilot survey); A/2014A/25,
A/2015A/19, A2017A/18 (The GALAH survey phase 1); A2018A/18 (Open clusters with HERMES);
A2019A/1 (Hierarchical star formation in Ori OB1); A2019A/15, A/2020B/23, R/2022B/5,
R/2023A/4, R2023B/5 (The GALAH survey phase 2); A/2015B/19, A/2016A/22, A/2016B/10,
A/2017B/16, A/2018B/15 (The HERMES-TESS program); A/2015A/3, A/2015B/1, A/2015B/19,
A/2016A/22, A/2016B/12, A/2017A/14, A/2020B/14 (The HERMES K2-follow-up program);
R/2022B/02 and A/2023A/09 (Combining asteroseismology and spectroscopy in K2);
A/2023A/8 (Resolving the chemical fingerprints of Milky Way mergers); and A/2023B/4
($s$-process variations in southern globular clusters). We acknowledge the traditional
owners of the land on which the AAT stands, the Gamilaraay people, and pay our respects
to elders past and present. This paper includes data that has been provided by AAO Data
Central (datacentral.org.au). This work has made use of data from the European Space
Agency (ESA) mission Gaia (https://www.cosmos.esa.int/gaia), processed by the Gaia Data
Processing and Analysis Consortium (DPAC, https://www.cosmos.esa.int/web/gaia/dpac/consortium).
Funding for the DPAC has been provided by national institutions, in particular the
institutions participating in the Gaia Multilateral Agreement.
\end{acknowledgments}


\begin{thebibliography}{99}
\bibitem[Abbott et al.(2017)]{Abbott2017} Abbott, B. P., Abbott, R., Abbott, T. D., et al. 2017, ApJL, 850, L39
\bibitem[Allen \& Barbuy(2006a)]{Allen2006a} Allen, D. M., \& Barbuy, B. 2006a, A\&A, 454, 895
\bibitem[Allen \& Barbuy(2006b)]{Allen2006b} Allen, D. M., \& Barbuy, B. 2006b, A\&A, 454, 917
\bibitem[Arcones \& Thielemann(2023)]{Arcones2023} Arcones, A., \& Thielemann, F. K. 2023, A\&ARv, 31, 1A
\bibitem[Asplund et al.(2009)]{Asplund2009} Asplund, M., Grevesse, N., Sauval, A. J., \& Scott, P., 2009, ARA\&A, 47, 481
\bibitem[Astropy Collaboration et al.(2013)]{Astropy2013} Astropy Collaboration, Robitaille, T. P., Tollerud, E. J., et al. 2013, A\&A, 558, A33
\bibitem[Astropy Collaboration et al.(2018)]{Astropy2018} Astropy Collaboration, Price-Whelan, A. M., Sip\'{o}cz, B. M., et al. 2018, AJ, 156, 123
\bibitem[Astropy Collaboration et al.(2022)]{Astropy2022} Astropy Collaboration, Price-Whelan, A. M., Lim, P. L., et al. 2022, ApJ, 935, 167
\bibitem[Bayo et al.(2008)]{Bayo2008} Bayo, A., Rodrigo, C., \& Barrado Y Navascu\'{e}s, D. 2008, A\&A, 492, 277
\bibitem[Beers \& Christlieb(2005)]{Beers2005} Beers, T. C., \& Christlieb, N. 2005, ARA\&A, 43, 531
\bibitem[Belokurov et al.(2018)]{Belokurov2018} Belokurov, V., Erkal, D., Evans, N. W., et al. 2018, MNRAS, 478, 611
\bibitem[Belokurov et al.(2020)]{Belokurov2020} Belokurov, V., Sanders, J. L., Fattahi, A., et al. 2020, MNRAS, 494, 3880
\bibitem[Bennett \& Bovy(2019)]{Bennett2019} Bennett, M., \& Bovy, J. 2019, MNRAS, 482, 1417
\bibitem[Bianchi et al.(2017)]{Bianchi2017} Bianchi, L., Shiao, B., \& Thilker, D. 2017, ApJS, 230, 24
\bibitem[Bidelman \& Keenan(1951)]{Bidelman1951} Bidelman, W. P., \& Keenan, P. C. 1951, ApJ, 114, 473
\bibitem[Bisterzo et al.(2012)]{Bisterzo2012} Bisterzo, S., Gallino, R., Straniero, O., et al., 2012, MNRAS, 422, 849
\bibitem[B\"{o}hm-Vitense et al.(2000)]{Vitense2000} B\"{o}hm-Vitense, E., Carpenter, K., Robinson, R., et al. 2000, ApJ, 533, 969
\bibitem[Brusadin et al.(2013)]{Brusadin2013} Brusadin, G., Matteucci, F., \& Romano, D. 2013, A\&A, 554, A135
\bibitem[Bunder et al.(2024)]{Bunder2024} Bunder, S., Kos, J., Wang, E. X., et al. 2024, PASA, 42, e051
\bibitem[Burbidge et al.(1957)]{Burbidge1957} Burbidge, E. M., Burbidge, G. R., Fowler, W. A., \& Hoyle, F. 1957, RvMP, 29, 547
\bibitem[Burris et al.(2000)]{Burris2000} Burris, D. L., Pilachowski, C. A., Armandroff, T. E., et al. 2000, ApJ, 544, 302
\bibitem[Busso et al.(1999)]{Busso1999} Busso, M., Gallino, R., \& Wasserburg, G. J. 1999, ARA\&A, 37, 239
\bibitem[Busso et al.(2001)]{Busso2001} Busso M., Gallino R., Lambert D. L., Travaglio T., Smith V. V. 2001, ApJ, 557, 802  
\bibitem[Carrillo et al.(2024)]{Carrillo2024} Carrillo, A., Deason, A. J., Fattahi, A., et al. 2024, MNRAS, 527, 2165
\bibitem[Castelli \& Kurucz(2004)]{Castelli2004} Castelli, F., \& Kurucz, R. L. 2004, IAUS, 210, A20
\bibitem[Chen et al.(2004)]{Chen2004} Chen, Y. Q., Nissen, P. E., \& Zhao, G. 2004, A\&A, 425, 697
\bibitem[Chen et al.(2025)]{Chen2025} Chen, T. Y., Shi, J. R., Yan, H. L. 2025, ApJ, 982, 182
\bibitem[Cristallo et al.(2011)]{Cristallo2011} Cristallo, S., Piersanti, L., Straniero, O., et al. 2011, ApJS, 197, 17
\bibitem[Cseh et al.(2018)]{Cseh2018} Cseh, B., Lugaro, M., D'Orazi, V., et al. 2018, A\&A, 620, A146
\bibitem[Cseh et al.(2022)]{Cseh2022} Cseh, B., Vil\'{a}gos, B., Roriz, M. P., et al., 2022, A\&A, 660, A128
\bibitem[Cui et al.(2012)]{Cui2012} Cui, X. Q., Zhao, Y. H., Chu, Y. Q., et al. 2012, RAA, 12, 1197
\bibitem[de Castro et al.(2016)]{deCastro2016} de Castro, D. B., Pereira, C. B., Roig, F., et al. 2016, MNRAS, 459, 4299
\bibitem[De Silva et al.(2015)]{DeSilva2015} De Silva, G. M., Freeman, K. C., Bland-Hawthorn, J., et al. 2015, MNRAS, 449, 2604
\bibitem[Dodd et al.(2023)]{Dodd2023} Dodd, E., Callingham, T. M., Helmi, A., et al. 2023, A\&A, 670, L2
\bibitem[Escorza et al.(2017)]{Escorza2017} Escorza, A., Boffin, H. M. J., Jorissen, A., et al. 2017, A\&A, 608, A100
\bibitem[Escorza et al.(2019)]{Escorza2019} Escorza, A., Karinkuzhi, D., Jorissen, A., et al. 2019, A\&A, 626, A128
\bibitem[Escorza et al.(2020)]{Escorza2020} Escorza, A., Siess, L., Van Winckel, H., \& Jorissen, A. 2020, A\&A, 639, A24
\bibitem[Gaia Collaboration et al.(2023)]{Gaia2023} Gaia Collaboration, Recio-Blanco, A., Kordopatis, G., et al. 2023, A\&A, 674, A38
\bibitem[Gallino et al.(1998)]{Gallino1998} Gallino, R., Arlandini, C., Busso, M., et al. 1998, ApJ, 497, 388
\bibitem[Gilmore et al.(2002)]{Gilmore2002} Gilmore, G., Wyse, R. F. G., \& Norris, J. E. 2002, ApJ, 574, L39
\bibitem[G\'{o}mez et al.(1997)]{Gomez1997} G\'{o}mez A. E., Luri X., Grenier S., et al. 1997, A\&A, 319, 881
\bibitem[Goriely(1999)]{Goriely1999} Goriely, S. 1999, A\&A, 342, 881
\bibitem[Goswami \& Goswami(2023)]{Goswami2023} Goswami, P. P., \& Goswami, A. 2023, AJ, 165, 154
\bibitem[Gray \& Corbally(1994)]{Gray1994} Gray, R. O., \& Corbally, C. J. 1994, AJ, 107, 742
\bibitem[Han et al.(1995)]{Han1995} Han, Z. W., Eggleton, P. P., Podsiadlowski, P., \& Tout, C. A. 1995, MNRAS, 277, 1443
\bibitem[Han et al.(2022)]{Han2022} Han, J. J., Conroy, C., Johnson, B. D., et al. 2022, AJ, 164, 249
\bibitem[Harris et al.(2020)]{Harris2020} Harris, C. R., Millman, K. J., van der Walt, S. J., et al. 2020, Nature, 585, 357
\bibitem[Helmi et al.(2018)]{Helmi2018} Helmi, A., Babusiaux, C., Koppelman, H. H., et al. 2018, Nature, 563, 85
\bibitem[Horta et al.(2023)]{Horta2023} Horta, D., Schiavon, R. P., Mackereth, J. T., et al. 2023, MNRAS, 520, 5671
\bibitem[Hunter(2007)]{Hunter2007} Hunter, J. D. 2007, CSE, 9, 90
\bibitem[Johnson \& Soderblom(1987)]{Johnson1987} Johnson, D. R. H., \& Soderblom, D. R. 1987, AJ, 93, 864
\bibitem[Jorissen et al.(1998)]{Jorissen1998} Jorissen, A., Van Eck, S., Mayor, M., \& Udry, S. 1998, A\&A, 332, 877
\bibitem[Jorissen et al.(2019)]{Jorissen2019} Jorissen A., Boffin H. M. J., Karinkuzhi D., et al. 2019, A\&A, 626, A127
\bibitem[K\"{a}ppeler et al.(2011)]{kappeler2011} K\"{a}ppeler, F., Gallino, R., Bisterzo, S., \& Aoki, W. 2011, RvMP, 83, 157
\bibitem[Karinkuzhi et al.(2018)]{Karinkuzhi2018} Karinkuzhi, D., Van Eck, S., Jorissen, A., et al. 2018, A\&A, 618, A32
\bibitem[Karinkuzhi et al.(2021a)] {Karinkuzhi2021a} Karinkuzhi, D., Van Eck, S., Goriely, S., et al. 2021, A\&A, 645, A61
\bibitem[Karinkuzhi et al.(2021b)] {Karinkuzhi2021b} Karinkuzhi, D., Van Eck, S., Jorissen, A., et al. 2021, A\&A, 654, A140
\bibitem[Kluyver et al.(2016)]{Kluyver2016} Kluyver, T., Ragan-Kelley, B., P\'{e}rez, F., et al. 2016, in Loizides F., Schmidt B., eds
\bibitem[Kong et al.(2018)]{Kong2018} Kong, X. M., Kumar, Y. B., Zhao, G., et al. 2018, MNRAS, 474, 2129
\bibitem[Laporte et al.(2018)]{Laporte2018} Laporte, C. F. P., G\'{o}mez, F. A., Besla, G., et al. 2018, MNRAS, 473, 1218
\bibitem[Lebzelter et al.(2013)]{Lebzelter2013} Lebzelter, T., Uttenthaler, S., Straniero, O., \& Aringer, B. 2013, A\&A, 554, A30
\bibitem[Liang et al.(2000)]{Liang2000} Liang, Y. C., Zhao, G., \& Zhang, B. 2000, A\&A, 363, 555
\bibitem[Liang et al.(2003)]{Liang2003} Liang, Y. C., Zhao, G., Chen, Y. Q., et al. 2003, A\&A, 397, 257
\bibitem[Liu et al.(2009)]{Liu2009} Liu, G. Q., Liang, Y. C., \& Deng, L. C. 2009, RAA, 4, 431
\bibitem[Liu et al.(2021)]{Liu2021} Liu, S., Wang, L., Shi J. R., et al. 2021, RAA, 21, 278
\bibitem[Lu(1983)]{Lu1983} Lu, P. K., Dawson, D. W., Upgren, A. R., \& Weis, E. W. 1983, ApJS 52, 169
\bibitem[Lu(1991)]{Lu1991} Lu, P. K. 1991, AJ, 101, 2229
\bibitem[Majewski(2017)]{Majewski2017} Majewski, S. R., Schiavon, R. P., Frinchaboy, p. M., et al. 2017, AJ, 154, 94
\bibitem[Matteucci(2021)]{Matteucci2021} Matteucci, F. 2021, A\&AR, 29, 5
\bibitem[M\'{e}sz\'{a}ros \& Prieto(2013)]{Meszaros2013} M\'{e}sz\'{a}ros, S., \& Prieto, C. A. 2013, MNRAS, 430, 3285
\bibitem[McClure et al.(1980)]{McClure1980} McClure, R. D., Fletcher, J. M., \& Nemec, J. M. 1980, ApJ, 238, L35
\bibitem[McClure \& Woodsworth(1990)]{McClure1990} McClure, R. D., \& Woodsworth, A. W. 1990, ApJ, 352, 709
\bibitem[McMillan(2017)]{McMillan2017} McMillan, P. J. 2017, MNRAS, 465, 76
\bibitem[Naidu et al.(2020)]{Naidu2020} Naidu, R. P., Conroy, C., Bonaca, A., et al. 2020, ApJ, 901, 48
\bibitem[Norfolk et al.(2019)]{Norfolk2019} Norfolk, B. J., Casey, A. R., Karakas, A. I., et al. 2019, MNRAS, 490, 2219
\bibitem[Pereira et al.(2011)]{Pereira2011} Pereira, C. B., Sales Silva, J. V., Chavero, C., et al. 2011, A\&A, 533, A51
\bibitem[Pourbaix et al.(2004)]{Pourbaix2004} Pourbaix, D., Tokovinin, A. A., Batten, A. H., et al. 2004, A\&A, 424, 727
\bibitem[Price-Whelan(2017)]{Price2017} Price-Whelan, A. M. 2017, JOSS, 2, 388
\bibitem[Purandardas et al.(2019)]{Purandardas2019} Purandardas, M., Goswami, A., Goswami, P. P., et al. 2019, MNRAS, 486, 3266
\bibitem[Qian \& Wasserburg(2007)]{Qian2007} Qian, Y. Z., \& Wasserburg, G. J. 2007, PhR, 442, 237
\bibitem[Queiroz et al.(2023)]{Queiroz2023} Queiroz, A. B. A., Anders, F., Chiappini,  C., et al. 2023, A\&A, 673, A155
\bibitem[Rekhi et al.(2024)]{Rekhi2024} Rekhi, P., Sagi, B. A., Hallakoun, N., et al. 2024, ApJL, 973, L56
\bibitem[Rojas et al.(2013)]{Rojas2013} Rojas, M., Drake, N. A., Pereira, C. B., et al. 2013, Astrophysics, 56, 57
\bibitem[Roriz et al.(2021a)]{Roriz2021a} Roriz, M. P., Lugaro, M., Pereira, C. B., et al. 2021a, MNRAS, 501, 5834
\bibitem[Roriz et al.(2021b)]{Roriz2021b} Roriz, M. P., Lugaro, M., Pereira, C. B., et al. 2021b, MNRAS, 507, 1956
\bibitem[Roriz et al.(2024)]{Roriz2024} Roriz, M. P., Holanda, N., et al. 2024, AJ, 167, 184
\bibitem[Roriz et al.(2025)]{Roriz2025} Roriz, M. P., Drake, N. A., Holanda, N., et al. 2025, AJ, 170, 259
\bibitem[Ryu et al.(2025)]{Ryu2025} Ryu, T., Sari, R., de Mink, S. E., et al. 2025, A\&A, 702, A61
\bibitem[Sch\"{o}nrich et al.(2010)]{Schonrich2010} Sch\"{o}nrich, R., Binney, J., \& Dehnen, W. 2010, MNRAS, 403, 1829
\bibitem[Sheinis et al.(2015)]{Sheinis2015} Sheinis, A., Anguiano, B., Asplund, M., et al. 2015, Journal
of Astronomical Telescopes, Instruments, and Systems, 1, 035002
\bibitem[Shejeelammal et al.(2020)]{Shejeelammal2020} Shejeelammal, J., Goswami, A., Goswami, P. P., et al. 2020, MNRAS, 492, 3708
\bibitem[Shetye et al.(2020)]{Shetye2020} Shetye, S., Van Eck, S., Goriely, S., et al. 2020, A\&A, 635, L6
\bibitem[Sneden et al.(2008)]{Sneden2008} Sneden, C., Cowan, J. J., \& Gallino, R. 2008, ARA\&A, 46, 241
\bibitem[Stancliffe(2021)]{Stancliffe2021} Stancliffe, R. J. 2021, MNRAS, 505, 5554
\bibitem[Taylor(2005)]{Taylor2005} Taylor, M. B. 2005, ASPC, 347, 29T
\bibitem[Travaglio(1999)]{Travaglio1999} Travaglio, C., Galli, D., Gallino, R., et al. 1999, ApJ, 521, 691
\bibitem[Travaglio(2001)]{Travaglio2001} Travaglio, C., Burkert, A., \& Galli, D. 2001, NuPhA. A, 688, 396
\bibitem[Travaglio et al.(2004)]{Travaglio2004} Travaglio, C., Gallino, R., Arnone, E., et al. 2004, ApJ, 601, 864
\bibitem[Virtanen et al.(2020)]{Virtanen2020} Virtanen, P., Gommers, R., Oliphant, T. E., et al. 2020, NatMe, 17, 261
\bibitem[Warner(1965)]{Warner1965} Warner, B. 1965, MNRAS, 129, 263
\bibitem[Williams(1975)]{Williams1975} Williams, P. M. 1975, MNRAS, 170, 343
\bibitem[Yang et al.(2016)]{Yang2016} Yang, G. C., Liang, Y. C., Spite, M., et al. 2016, RAA, 16, 19
\bibitem[Yang et al.(2024)]{Yang2024} Yang, G. C., Zhao, J. K., Liang, Y. C., et al. 2024, MNRAS, 534, 3104
\bibitem[Yong et al.(2013)]{Yong2013} Yong, D., Norris, J. E., Bessell, M. S., et al. 2013, ApJ, 762, 26
\bibitem[Za\v{c}s(1994)]{Zacs1994} Za\v{c}s, L. 1994, A\&A, 283, 937
\bibitem[Zhao(2012)]{Zhao2012} Zhao, G., Zhao, Y.H., Chu, Y. Q., et al. 2012, RAA, 12, 723
\end{thebibliography}
\end{document}